\documentclass[11pt]{article}   	
\usepackage{geometry}                		
\usepackage[parfill]{parskip}    		

\usepackage{graphicx,color}				
\usepackage{amssymb}
\usepackage{hyperref}
\usepackage{authblk}

\title{Comparative analysis of SN1987A antineutrino fluence}
\author{Francesco Vissani}
\affil{\small Gran Sasso Science Institute (INFN), L'Aquila, Italy\\ 
Laboratori Nazionali del Gran Sasso (INFN),  Assergi (AQ), Italy}
\date{}

\begin{document}

\maketitle
\begin{abstract}
We discuss the electron antineutrino fluence 
derived from the events detected  by  Kamiokande-II, IMB and Baksan on February 23, 1987.
The data are analysed adopting a new simple and accurate formula for the signal,  improving on the previous modeling of the detectors response, considering the possibility of  background events. We perform  several alternative analyses  to quantify the relevance of various descriptions, approximations and biases. 
In particular, we study the effect of: omitting Baksan data or neglecting the background, using simplified formulae for the signal,  modifying the fluence to account for oscillations and pinching,   including the measured times and angles of the events,  using  other descriptions of detector response, etc. We show that most of these effects are small or negligible and argue, by comparing the allowed regions for astrophysical parameters, that the results are stable. We comment on the accordance with theoretical results and on open questions. 
\end{abstract}

\parskip3mm

\section{Introduction and motivations}

The events observed in underground neutrino detectors a
few hours before  SN1987A explosion are still the only direct validation of our understanding of what happens in the first instants after a gravitational collapse. As such, they are enormously important and after a quarter of a century they are still relevant for many scientific disciplines  including (neutrino-) astronomy, astrophysics, nuclear physics and particle physics. 

For these reasons the observations  
of Kamiokande-II \cite{kam} (Nobel 2002), IMB \cite{imb} and  Baksan~\cite{bak} have been studied several times over the years, applying a variety of methods and procedures. 

In this paper, we compare the methods of  analysis and the 
corresponding results,  aiming to assess their accuracy and to define  a  reference methodology that is as simple but also as reliable as possible. (Note that similar approaches are important whenever the sample of events is small.)

Discussions from scientific literature called attention 
on few general points 
\begin{itemize}
\item the hypothesis that the signal is due to electron antineutrinos is reliable
-- see \cite{haxton,olga} for critical discussion and note that its detection cross section is precisely known \cite{beacom,viss};
\item simple models should be preferred for the analysis, accounting for astrophysical uncertainties \cite{bahcall,bahcallBook,jeger}; 
\item there is no good reason to omit Baksan data, once the background is taken into account properly \cite{ll};
\item the LSD/Mont Blanc data \cite{lsd} have to be 
tentatively attributed to a speculative emission phase that preceded the main one by several hours \cite{deruju,bcg,olga} or,  conservatively, 
discarded from the analysis \cite{bahcall}.
\end{itemize}
In this paper, we discuss and quantify the relevance of these and many other  considerations for the analysis of the observed events.

\begin{table}[t]
\centerline{
\begin{tabular}{||c|rrcc||c|rrcc||}
\hline
& \footnotesize Relative\, & \footnotesize Energy &\footnotesize  SN-angle &\footnotesize  Backgr. & &\footnotesize Relative\, & \footnotesize Energy &\footnotesize  SN-angle &\footnotesize  Backgr. \\[-1ex]
&\footnotesize time \tiny [ms]  & \tiny [MeV]\ \ \ &\tiny  [deg] &\tiny  [Hz/MeV] & &\footnotesize time \tiny [ms]  & \tiny [MeV]\ \ \ &\tiny  [deg] &\tiny  [Hz/MeV]  \\ \hline
K1 & 0 & 20.0$\pm$2.9 &18$\pm$18 & 1.0E-5 & I1 & 0 & 38$\pm$7 &80$\pm$10 & $10^{-5}$?\\
K2 & 107 & 13.5$\pm$3.2 &40$\pm$27 & 5.4E-4 &I2 & 412 & 37$\pm$7 &44$\pm$15 & $10^{-5}$?\\
K3 & 303 & 7.5$\pm$2.0 &108$\pm$32 & 2.4E-2 &I3 & 650 & 28$\pm$6 &56$\pm$20 & $10^{-5}$?\\
K4 & 324 & 9.2$\pm$2.7 &70$\pm$30 &2.8E-3&I4 & 1141 & 39$\pm$7 &65$\pm$20 &$10^{-5}$?\\
K5 & 507 & 12.8$\pm$2.9 &135$\pm$23 & 5.3E-4&I5 & 1562 & 36$\pm$9 & 33$\pm$15& $10^{-5}$?\\
K6 & 686& 6.3$\pm$1.7 & 68$\pm$77& 7.9E-2&I6 & 2684& 36$\pm$6 &52$\pm$10 & $10^{-5}$?\\
K7 & 1541& 35.4$\pm$8.0 &32$\pm$16 & 5.0E-6&I7 & 5010& 19$\pm$5 &42$\pm$20 & $10^{-5}$?\\
K8 & 1728& 21.0$\pm$4.2 &30$\pm$18 & 1.0E-5&I8 & 5582& 22$\pm$5 & 104$\pm$20& $10^{-5}$?\\ 
K9 & 1915& 19.8$\pm$3.2&38$\pm$22 & 1.0E-5&&&&&\\
K10 & 9219& 8.6$\pm$2.7&122$\pm$30 & 4.2E-3 &&&&&\\
K11 & 10433& 13.0$\pm$2.6&49$\pm$26 & 4.0E-4&&&&&\\
K12 & 12439& 8.9$\pm$2.9 &91$\pm$39 & 3.2E-3&B1 & 0 & 12.0$\pm$2.4 & 90? & 8.4E-4\\
K13 & 17641 & 6.5$\pm$1.6& 90? & 7.3E-2&B2 & 435 & 17.9$\pm$3.6 & 90? & 1.3E-3\\
K14 & 20257&  5.4$\pm$1.4& 90? &5.3E-2&B3 & 1710 & 23.5$\pm$4.7 & 90? & 1.2E-3\\
K15 & 21355& 4.6$\pm$1.3& 90? & 1.8E-2&B4 & 7687 & 17.5$\pm$3.5 & 90? & 1.3E-3\\ 
K16 & 23814& 6.5$\pm$1.6& 90? &7.3E-2  & B5 & 9099 & 20.3$\pm$4.1 & 90? & 1.3E-3\\ \hline
\end{tabular}}
\caption{\em \small Properties of the events in the neutrino bursts observed  in the occasion of SN1987A. The events are indicated by K1, K2... K16 for Kamiokande-II; I1, I2... I8 for IMB; B1, B2 ... B5 for Baksan. 
The  background rate of IMB and some of the angles  are unknown;  
there is no  harm (for the analysis) in setting their values as indicated in the table with the question marks. \label{tab0}}
\end{table}

We use the energy spectrum of the events as the basis for comparison;  
moreover, we focus on the simplest description of the time integrated flux (=fluence) of electron antineutrinos, namely
\begin{equation}\label{flenza}
\frac{dF}{dE_\nu}=\frac{\mathcal{E}}{4\pi D^2 } \times \frac{E_\nu^2 \ e^{-E_\nu/T}}{6\ T^4}
\end{equation}
where $E_\nu$ is the electron antineutrino energy and 
$D$ is the distance from the supernova, that we assume to be 
50 kpc -- its uncertainty being 
of the order of 5\%.
The fluence  parameters,  
$\mathcal{E}$ and $T$, describe respectively:
\begin{enumerate}
\item The amount of   energy radiated in electron antineutrinos.  
The formation of a compact stellar remnants requires that  
 $10-20$\% of the rest core's mass is radiated in neutrinos and antineutrinos of all species.
The fraction carried away by each of the six species, 
according to the numerical simulations of the supernova,
should be about the same, i.e.\ 1/6.\footnote{Although this property is occasionally called `equipartition',  to the best of our knowledge it has no fundamental meaning and in fact, it is expected to be obeyed up to a factor of 2. Note also that the total energy  is 
$2-3$ order of magnitude larger than the kinetic energy of the explosion.} For these reasons, we expect that $\mathcal{E}$ lies in 
the few $10^{52}$ erg range.
\item The temperature of the electron antineutrinos. This is simply connected to the average energy as 
\begin{equation}\label{smpq}\langle E_{\nu}\rangle=3 T
\end{equation} 
The value of $T$, expected to be in the few MeV range,
is dictated by the interactions of the neutrinos
and by the distribution of the matter around 
the last scattering surface of the forming 
compact stellar object.
\end{enumerate}
Thanks to Eq.~\ref{flenza}, the 
comparison with earlier reference 
works as \cite{bahcall,bahcallBook,jeger}
 is easy 
 and the 
 output parameters have immediate physical meaning: e.g.\ the total fluence of $\bar\nu_e$ is given by the analytical expression  
$F=\mathcal{E}/(4\pi D^2 3 T)$. 
 The hypothesis described by Eq.~\ref{flenza} will be assessed 
 later on in this paper.

The outline of this investigation is as follows. We propose a methodology and we use it to perform a detailed study of the electron antineutrinos emitted from SN1987A; 
then, we  compare the result with those of several alternative approaches. In Sect.~\ref{pelos}  we clarify 
certain technical points needed for the analysis--in particular, those regarding the  description of the detector (Sect.~\ref{forma}) and  the cross section (Sect.~\ref{sibd}). Then, we begin with an accurate  analysis of the individual data sets, arguing that the spectra  are compatible, and obtaining the parameters $\mathcal{E}$ and $T$ by means of a combined analysis of all the data (Sect.~\ref{ra}). Finally, we compare the result of this accurate analysis with those of other analyses in Sect.~\ref{aa}, quantifying the impact of alternative assumptions and/or of alternative procedures in
Sect.~\ref{cc}. The results are discussed in Sect.~\ref{culdis}
and summarized in Sect.~\ref{o1}.

\begin{figure}[t]
\centerline{\includegraphics[width=0.6\linewidth]{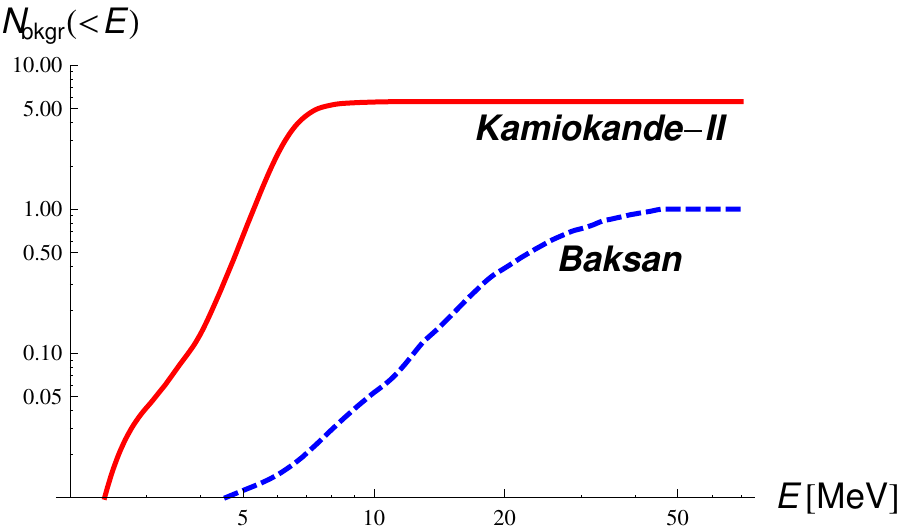}}
    \caption{\em \small Cumulative distribution of the expected number of background events in a time window of 30 seconds,
  as a function of the visible energy. IMB's curve is 
  assumed to be lower in this scale.
  \label{fx}
  }
  \label{fig:buco}
\end{figure}

\section{Technical details\label{pelos}}

The available data are given in Tab.~\ref{tab0}. 
The information from the experimental collaborations is taken from \cite{kam,imb,bak}.
Note that we list in Tab.~\ref{tab0} all events in a wide time and energy window, without  {\em a priori} assuming
that some of them are due to signal and some other to background.  

In this section, we describe the likelihood we adopt  
and then discuss  the response of the detectors to the signal. An accurate 
formalism for the analysis of these data has been described in 
Jegerlehner, Neubig and Raffelt 
\cite{jeger},  and we review it in  Sect.~\ref{purea}.
However, this formalism 
has not yet been applied to the detailed type of analysis we adopt here.
Also note that the formula we used here to describe the signal is a new, improved approximation of the numerical formula that is usually adopted (compare  Eqs.~\ref{megl} and \ref{tred}). It is accurate and particularly well-suited to numerical analysis.

\subsection{General framework}
In this subsection, we focus on the general aspects, namely: 
the type of likelihood, the hypotheses on the background, 
the formalism to describe detectors,  our assumptions on the signal. We will discuss in the subsequent two sections the new specific  implementations that we use in our analysis.
\subsubsection{Poissonian likelihood}
The most efficient statistical tool to analyze small data sets is the un-binned (Poissonian) likelihood, see e.g.\ \cite{fernando}. For any individual detector, this is,
\begin{equation}\label{lugu}
\mathcal{L}(x)= e^{-N_{\mbox{\tiny tot}}(x) }\times \prod_{i=1}^{N_{\mbox{\tiny obs}}} dN_i(x)
\end{equation}
where the number of observed events is $N_{\mbox{\tiny obs}}$, 
and $x$ are the model-parameters that we can estimate by maximizing the likelihood. Focussing on  the  study of the energy distribution (spectrum), the 
expected number of events around the 
{\em observed energy} $E_i$ is given by 
 \begin{equation}\label{above}
 dN_i=dE \left[ \frac{dS}{dE_i} + \frac{dB}{dE_i} \right]
 \end{equation}
 where the first term is due to signal (that we discuss later on), the second to
 background (that  we discuss in the subsequent section), and the width $dE$ is small enough to contain at most one 
 event.\footnote{Note that, being a constant, the actual value of $dE$ is irrelevant for the statistical tests we perform here as they are based on likelihood {\em ratios}.} Finally, the integral of $dN_i$ gives the total number of expected events $N_{\mbox{\tiny tot}}$. Assuming the background is zero or setting it to its known value we can perform inferences on the neutrino emission. In fact the expectation of the signal depends upon free parameters which are determined by the likelihood analysis. 
 
 \subsubsection{Background}
The background counting rate  can be measured very precisely whenever we deal with a stable detector. Kamiokande-II collaboration \cite{kam} published a complete information on their counting rate.  Ref.~\cite{ll} collected a great deal of information useful and relevant for the analysis, including the data regarding the background counting rate in Baksan. The available information (that we assume) 
is summarized in Tab.~\ref{tab0} and Fig.~\ref{fig:buco}. 

Note that we adopt the same background counting rate as given in Fig.~2 of~\cite{ll}. We do not perform the further convolution with the response function recommended in \cite{ll}, since the figure gives directly the measured value of the background rate, 
as argued in \cite{jcap,pagl}; see~\cite{jcap} for further discussion concerning the use of the background distribution and cross checks, using the information provided by Kamiokande-II.

 In our analysis, based on Eq.~\ref{above}, we use the values of  $dB/dE_i$ 
 as given in Tab.~\ref{tab0}, and multiplied by the pre-selected time of duration of the analysis. By default, and if not specified otherwise, this is of 30 seconds, that allows us to include all events of Tab.~\ref{tab0}. The total number of background events can be read from Fig.~\ref{fig:buco} and from the fourth column of Tab.~\ref{tab1}.

\subsubsection{Linear response of the detectors\label{purea}}

We describe the observed spectrum of the signal by assuming
a linear response of the detector \cite{jeger,fernando}. In this case, we  need to know the 
{\em intrinsic efficiency function} $\eta(E_e)$ and the 
{\em smearing function} $G$;
\begin{equation}\label{petella}
\frac{dS}{dE_i}=\int_{0}^\infty  \eta(E_e)\,  G\Big(E_e-E_i\, ,\, \sigma(E_e)\Big)\ \frac{dS_e}{dE_e}\ dE_e
\end{equation}
where $S$ is the observed distribution and 
$S_e$ the true distribution due to signal, while 
$E_i$ is the observed energy and $E_e$ the true one. 
The distribution of observed (reconstructed) positron energy $E$, for a given  true value of the energy $E_e$, satisfies 
$\int_{-\infty}^\infty G(E_e-E,\sigma)\ dE=1$, where the lower limit can be replaced by 0 as an excellent approximation in our case. We assume that the smearing function  
has a  Gaussian form, 
namely 
\begin{equation}\label{gusgus}
G(x,\sigma)={e^{-{x^2}/ (2 \sigma^2)}}/(\sqrt{2 \pi}\sigma)\end{equation} and discuss the explicit form of the 
{\em uncertainty function} $\sigma$ later on.

By calculating the total number of events above a certain minimum observed energy (threshold of analysis or in short threshold) 
$E>E_{\mbox{\tiny min}}$, 
we find 
\begin{equation}S(E_{\mbox{\tiny min}})=\int_{0}^\infty  \epsilon(E_e,E_{\mbox{\tiny min}})\ \ \frac{dS_e}{dE_e}\ dE_e
\label{tuttic}\end{equation} 
where the 
{\em total 
efficiency} is defined to be
\begin{equation}\label{meld}
\epsilon(E_e,E_{\mbox{\tiny min}})=\eta(E_e)\times g(E_e,E_{\mbox{\tiny min}})\mbox{ with }
 g(E_e,E_{\mbox{\tiny min}})=\frac{1+\mbox{Erf}\left[ \frac{E_e- E_{\mbox{\tiny min}} }{\sqrt{2} \sigma(E_e)}\right]}{2}
\end{equation}
and where Erf is defined as $\int_0^y G(x,\sigma) dx=\mbox{Erf}[\, y/(\sqrt{2}\sigma) \, ]/2  $.

\begin{figure}[t]
\centerline{\includegraphics[width=0.7\linewidth]{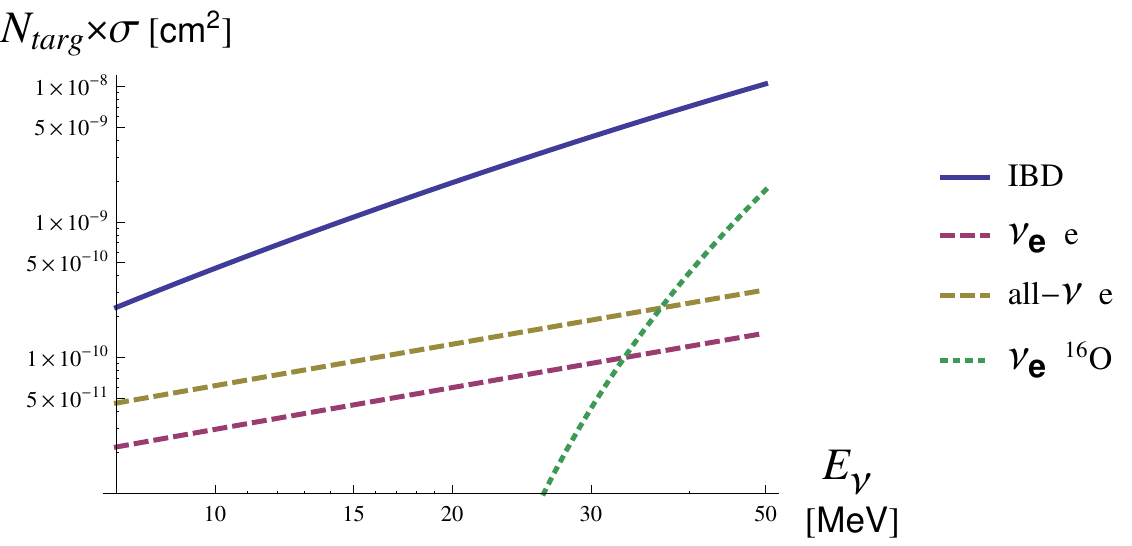}}
  \caption{\em \small Various neutrino cross section times the number of targets in 1 kton of water as a function of neutrino energy. 
  The ES cross section for $\nu_e$ scattering is compared with the one summed over all neutrinos and antineutrinos, 
  that is about twice larger.
  \label{fx}
  }
  \label{fig:csec}
\end{figure}

\subsubsection{Interactions}

As we have already mentioned it and as illustrated in Fig.~\ref{fig:csec}, the inverse beta decay  (IBD)
\begin{equation}\label{fib}
\bar\nu_e+p\to e^++n
\end{equation}
is by far the largest at the relevant energies for supernova neutrinos.
The elastic scattering (ES) cross section is characterized by directional events, whereas the interaction of electron neutrinos $\nu_e$ on oxygen
\cite{haxton}
(Oxy) only happens to matter for neutrinos with pretty high energies.
However, the number of observable events from both reactions is further  suppressed because 
the ES reaction also produces a neutrino which carries the energy away and the Oxy reaction has a relatively high threshold, about 15 MeV. See \cite{prdML}, \cite{aanda} and \cite{tommaso} for 
further discussion. Most neutral current reactions give minor contributions because the cross section is small at the relevant energies, or the products of the reaction are not observable. An extreme case is  the elastic scattering on proton $\nu+ p\to \nu+ p$,  
where the scattered proton has a very low energy that prevents its observation.
Thus, in this work we will suppose that the observed signal events are due to the positrons emitted in the inverse beta decay reaction. 
(We will test this hypothesis  {\em a posteriori}: In  Sect.~\ref{q4} we consider neutral current events, potentially relevant for Baksan and for Kamiokande-II; in  Sect.~\ref{p42} we study 
the elastic scattering events on electron, potentially relevant for Kamiokande-II and IMB.)

In a water Cherenkov detector, as IMB or Kamiokande-II, a positron 
behaves just as an electron, and the above formalism to describe the signal stays unchanged; one simply has to set 
the energy threshold in Eqs.~\ref{petella} and \ref{tuttic} to the value 
desired for data analysis.
In a scintillator, things are slightly different; indeed, the visible energy of the electron is the kinetic energy
$T_e=E_e-m_e$, whereas a positron  will add observable energy, due to its annihilation. 
For this reasons, 
we described the response of the scintillator by
replacing 
\begin{equation}E_e\to E_e+ m_e\end{equation}
in the response function of the detector $\eta(E_e)\times G(E_e-E_i,\sigma(E_e))$
appearing in Eqs.~\ref{petella} and 
also in the total efficiency appearing in Eq.~\ref{tuttic}.

\begin{figure}[t]
\begin{center}
\centerline{\includegraphics[width=0.31\linewidth]{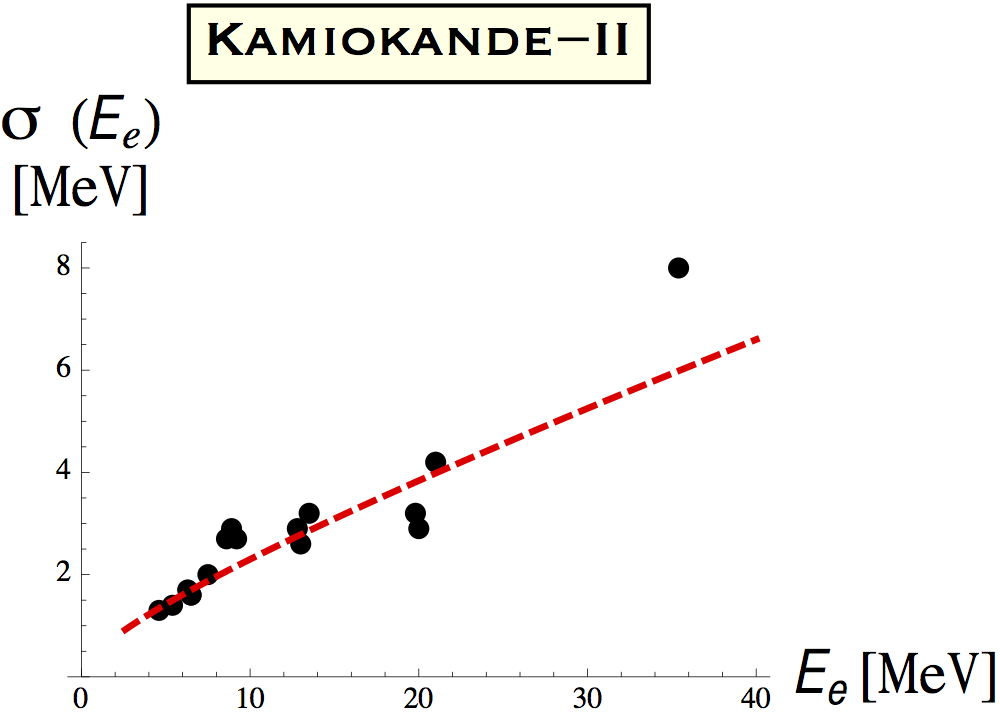}\hfil
\includegraphics[width=0.31\linewidth]{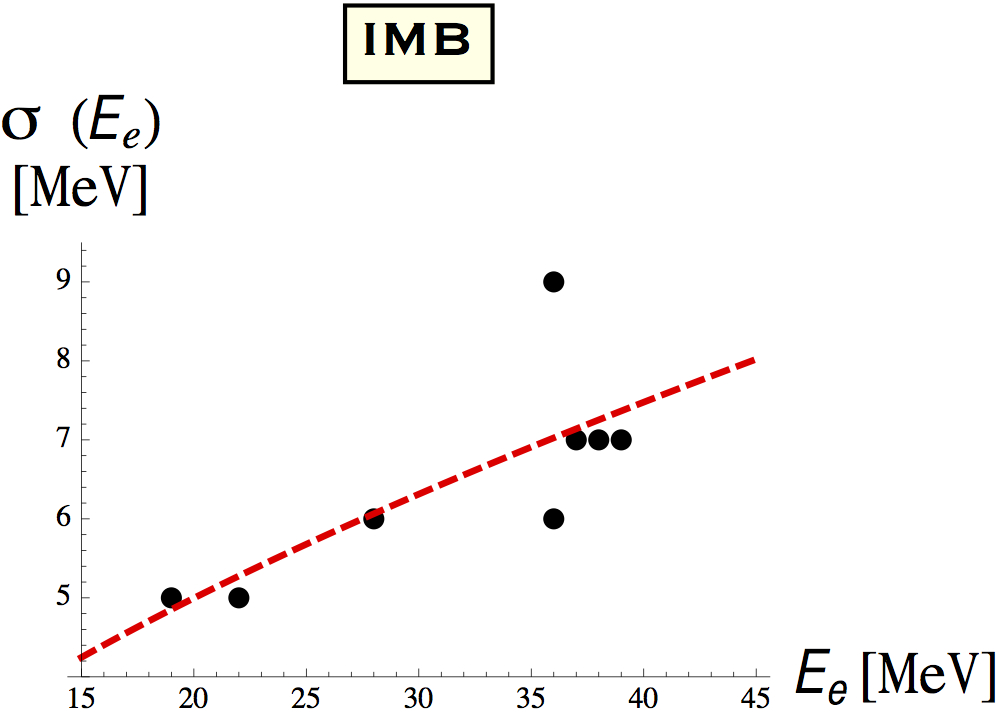}\hfil
\includegraphics[width=0.31\linewidth]{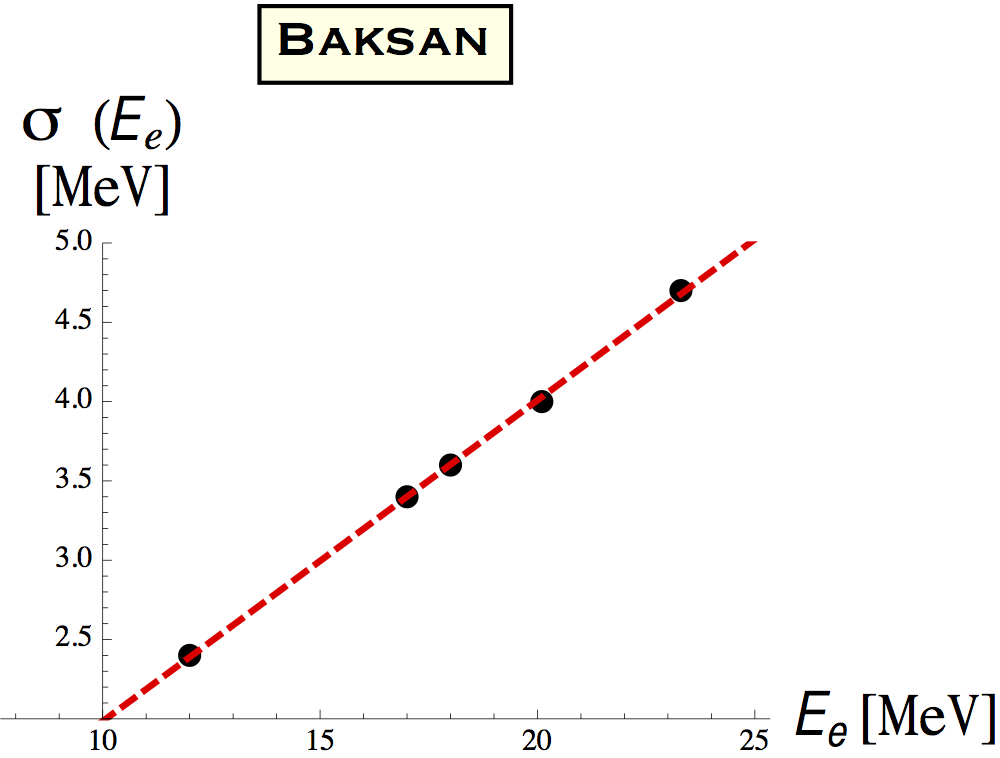}}
\caption{\em\small Experimental values of the error in the energy as a function of the energy of the event, along with the analytical description 
based on Eq.~\ref{erri} and Tab.~\ref{tab1}.}
\label{sigm}
\end{center}
\end{figure}

  \subsection{Detailed description of Kamiokande-II, IMB, Baksan \label{forma}}

We consider the statistical and the systematical components of the 
uncertainty function $\sigma(E_e)$ appearing in Eq.~\ref{gusgus}, i.e.,
\begin{equation}\label{erri}
\sigma(E_e)= \sigma_{\mbox{\tiny stat}} 
\times \left(\frac{E_e}{10\mbox{ MeV}}\right)^{1/2}+
\sigma_{\mbox{\tiny syst}} \times 
\left(\frac{E_e}{10\mbox{ MeV}}\right)
\end{equation}
We determine the two coefficients by using the energies of the observed events and their errors \cite{kam,imb,bak}. The result, reported in Tab.~\ref{tab1} and illustrated in Fig.~\ref{sigm}, 
was found consistent 
with the available information  on solar neutrinos at lower 
energies~\cite{suola}.

\begin{figure}[t]
\begin{center}
\centerline{\includegraphics[width=0.31\linewidth]{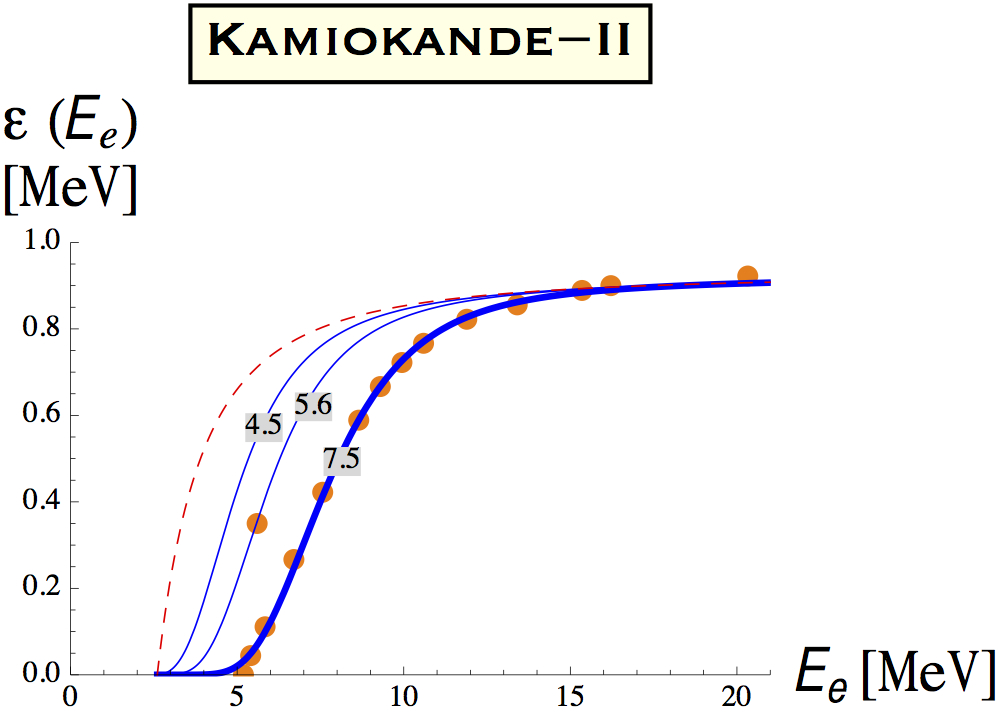}\hfil
\includegraphics[width=0.31\linewidth]{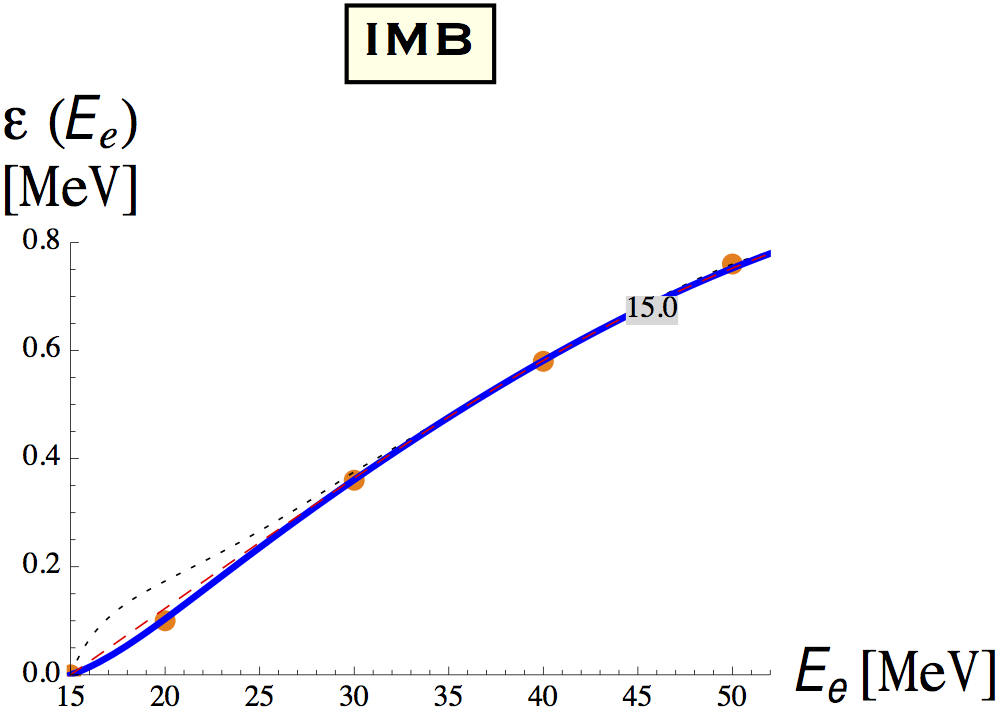}\hfil
\includegraphics[width=0.31\linewidth]{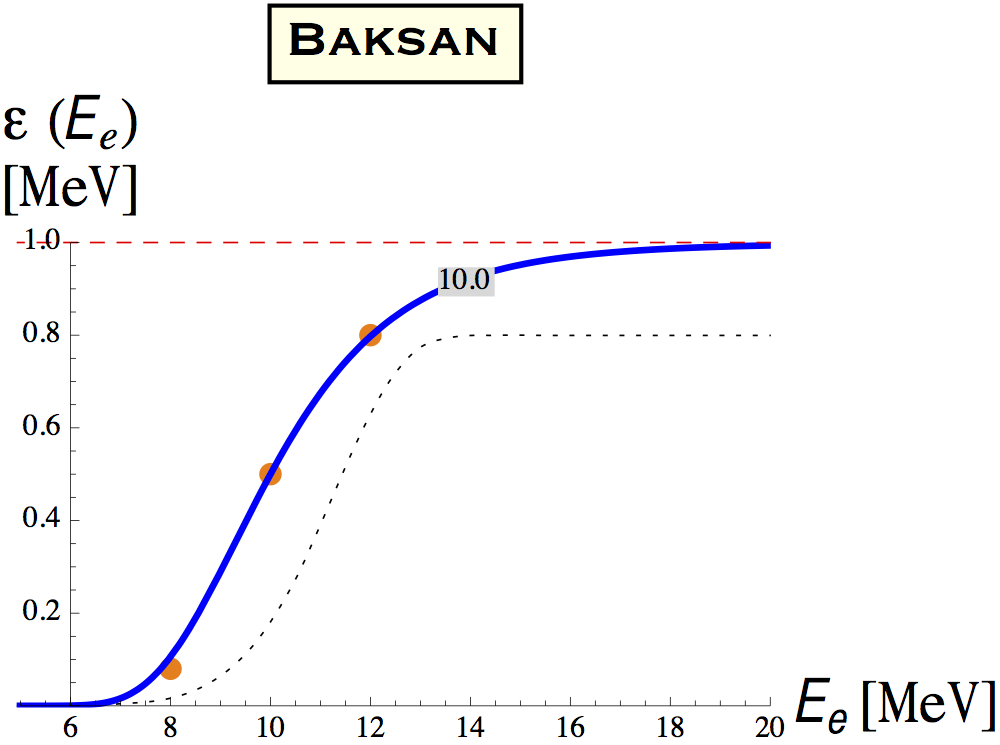}}
\caption{\em 
\small The orange dots are values of $\epsilon$
provided by the experimental collaborations.
Efficiency curves used in this work are plotted in blue for the choices of  
$E_{\mbox{\tiny min}}/\mbox{MeV}$ indicated there while the curves for $\eta$ (coinciding with the efficiency for $E_{\mbox{\tiny min}}=0$) 
 are given by dashed red lines. 
The two gray dotted curves are as follow: for IMB, this is the curve of 
$\eta$ used  by \cite{jeger} (the difference is mostly due to the different value of 
$E_{\mbox{\tiny min}}$); for Baksan, it is the efficiency curve used by \cite{ll} which does not reproduce the three experimental points very well.}
\label{effs}
\end{center}
\end{figure}

At this point, we can use  the 
published information on the total efficiency $\epsilon$ 
and on the threshold (again Tab.~\ref{tab1})
to extract the intrinsic efficiency from  
Eq.~\ref{meld}, written as,
\begin{equation}\eta(E_e)=\frac{\epsilon(E_e,E_{\mbox{\tiny min}})}{g(E_e,E_{\mbox{\tiny min}})}\end{equation}
Of course, this procedure becomes less reliable below threshold $E_{\mbox{\tiny min}}$, however we can assume that 
$\eta$ is a smooth, increasing function of the energy and we can perform some tests on the result.\footnote{Another procedure
is to use the published efficiency along with the scaling law
\begin{equation}
\epsilon(E_e,E_{\mbox{\tiny min}})=
\epsilon(E_e,7.5\mbox{ MeV}) \times \frac{g(E_e,E_{\mbox{\tiny min}})}{g(E_e,7.5\mbox{ MeV})}
\end{equation}
There is no real difference with our procedure, since the analytical description of $\eta(E_e)$ we propose turns out to provide a quite accurate description of the published efficiency, see Fig.~\ref{effs}.
Moreover, an extrapolation of $\epsilon$ to the limit
$E_{\mbox{\tiny min}} \to 0$, namely $\eta(E_e)$, is necessary for the 
statistical analysis.}

The case of Baksan \cite{bak} is the simplest; the function $\eta$ can be set to 1, since the number of photoelectrons collected is large. At the opposite extreme, there is the case of IMB \cite{imb}, where $\eta$ deviates strongly from 1, due to partial functioning. We have  obtained a very good description of its published efficiency by the following  parameterization of the intrinsic efficiency,
\begin{equation}\label{lgoggi}
\eta(E_e)=\sum_{n=1}^5 c_n(\mbox{\tiny IMB})\times \left(\frac{E_e}{E_{\mbox{\tiny IMB}}} - 1\right)^{\!n}
 \mbox{ for } E_{\mbox{\tiny IMB}}<E_e<70\mbox{ MeV}
\end{equation} 
with $c_1(\mbox{\tiny IMB})=0.369$, $c_4(\mbox{\tiny IMB})=-6\times 10^{-4}$,
$c_5(\mbox{\tiny IMB})=10^{-4}$ while $c_2(\mbox{\tiny IMB})=c_3(\mbox{\tiny IMB})=0$.
The uncertainties indicated in \cite{imb} allow to increase or decrease $\eta$ by 0.05 and 
can be very conveniently described by varying  the parameter $E_{\mbox{\tiny IMB}}$ (that could be thought of as an 
intrinsic, or hardware, threshold)
in the following range 
\begin{equation}\label{lpilo}E_{\mbox{\tiny IMB}}= 15\pm 2\mbox{ MeV}\end{equation}
Finally, the intrinsic efficiency of Kamiokande-II \cite{kam}
deviates from unity as well. This is due e.g.\ to geometrical effects: for instance, the events produced 
close to the walls are hardly detected and this is particularly true at low energy. 
This intrinsic efficiency can be approximated by
\begin{equation}
\eta(E_e)=0.93 \left[1-\left(\frac{0.2\mbox{ MeV}}{E_e}\right)-\left(\frac{2.5\mbox{ MeV}}{E_e}\right)^2 \right]\mbox{ for }E>E_{\mbox{\tiny KII}}=2.6\mbox{ MeV}
\end{equation}
With this expression, we find that  $\epsilon=35$\% when the analysis threshold is lowered to 
$E_e >E_{\mbox{\tiny min}}=5.6$ MeV, just as stated in \cite{kam} and as illustrated in Fig.~\ref{effs}, 
leftmost plot.

\begin{figure}[t]
\begin{center}
\centerline{\includegraphics[width=0.5\linewidth]{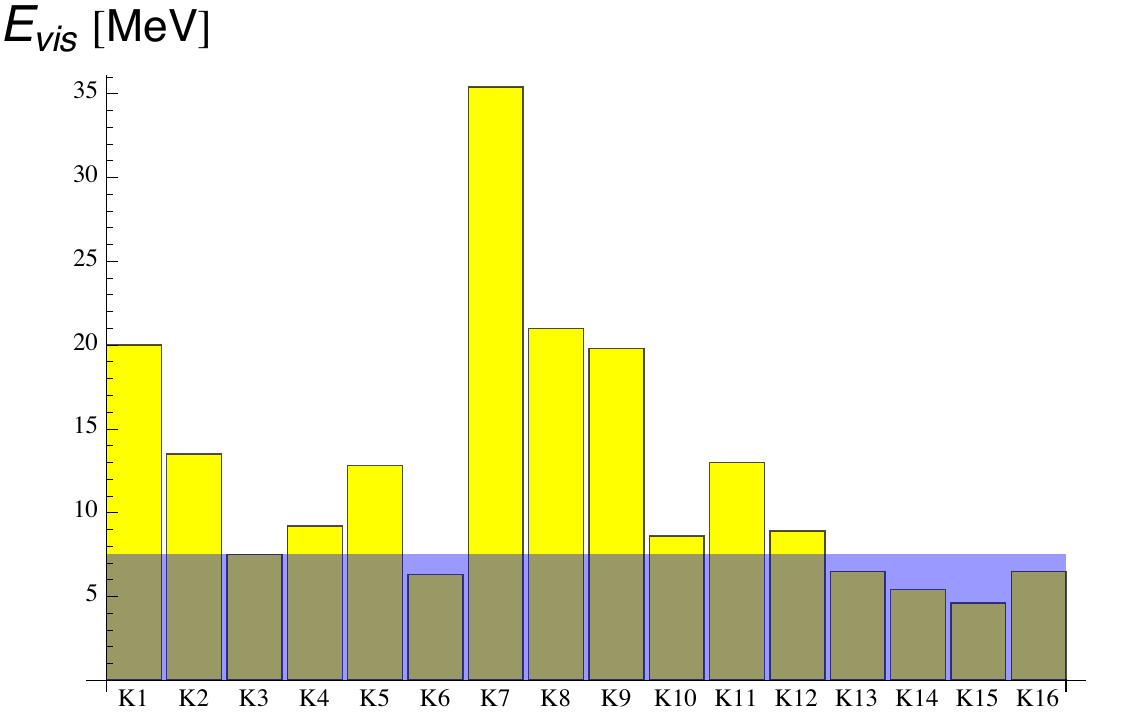}}
\caption{\em 
\small 
Progressive number and energies of the 16 events listed in Tab.~\ref{tab0} and observed by Kamiokande-II, to illustrate the assumption $E\ge 7.5$ MeV, that is often used to exclude {\em a priori}
the background events and to identify  the signal events.}
\label{pipin}
\end{center}
\end{figure}

Few remarks on these results are in order.
\begin{itemize}
\item The general formalism that we have described above has been outlined already in \cite{jeger} but its application to extend the analysis to the low energy events of SN1987A is new.
\item The above simple analytical forms (that are also new) are to be thought of as reasonable descriptions of the intrinsic efficiency and have no fundamental meaning. However, according to Fig.~\ref{effs} they reproduce well the published information. In addition, they are very convenient for the analysis. 
\item
Choosing a threshold of
$E_{\mbox{\tiny min}}=4.5$ MeV in Kamiokande-II, see Tab.~\ref{tab1}, allows us to include all events from Tab.~\ref{tab0}, with a minor extrapolation of the efficiency to lower energies.
We will use this threshold in the rest of this work, unless otherwise stated.
\end{itemize}
In order to illustrate better the meaning of the last assumption, we 
depict in Fig.~\ref{pipin} the usual criterion 
$E\ge E_{\mbox{\tiny min}}=7.5$ MeV, often adopted 
 to separate signal from background events in the dataset of Kamiokande-II. Note that in this manner the five events K6, K13, K14,K15, K16 are excluded {\em a priori} from the analysis, whereas K3 has exactly the value of the assumed energy threshold
and by definition it is kept. The alternative criterion 
$E\ge E_{\mbox{\tiny min}}=4.5$ MeV,
(that we adopt here unless stated otherwise) allows us to include all 16 events in the analysis and to 
 identify the background events 
 {\em a posteriori}, taking advantage of the entire available information.

\subsection{Modeling the signal from inverse beta decay\label{sibd}}

The distribution of the 
signal due to IBD, Eq.~\ref{fib},  
should be obtained from the integral over the allowed antineutrino energies,
 \begin{equation} \label{csec}
 \frac{dS_e}{dE_e}=N_p
 \int_{E_{\nu}^{\mbox{\tiny max}}}^{E_{\nu}^{\mbox{\tiny min}}} dE_\nu \ 
 \frac{dF}{dE_\nu}(E_\nu) \ 
 \frac{d\sigma_{\mbox{\tiny IBD}}}{dE_e}(E_\nu,E_e) 
 \end{equation}
 where $N_p$ is the number of target protons (see Tab.~\ref{tab1}), $F$ is the fluence
 (=time integrated flux) and $\sigma_{\mbox{\tiny IBD}}$ the cross section.
 Virtually exact expressions of the differential cross section  
 ${d\sigma}_{\mbox{\tiny IBD}}/{dE_e}$ are given in \cite{beacom,viss}. In the range of relevant energies, we can obtain the limits of integration by considering the cases $\cos\theta=\pm 1$ in the expression
\begin{equation}\label{midpoint}
E_\nu= \frac{E_e+\delta}{1-\frac{E_e-p_e\cos\theta}{m_p}}\mbox{ with }\delta=\frac{m_n^2-m_p^2-m_e^2}{2m_p}
\end{equation}
where $E_e$ and $p_e$ are the energy and momentum of the positron, $\theta$ the scattering angle with the direction of the antineutrino, and 
$m_p,m_n,m_e$ are the masses of the proton, neutron and electron respectively. Note, 
\begin{equation}\delta=1.294\mbox{ MeV }\approx \Delta=m_n-m_p=1.293\mbox{ MeV}
\end{equation}
so both quantities can be exchanged for all practical purposes.

\begin{table}[t]
 \centerline{
 \begin{tabular}{|c|ccccc|}
 \hline
& $N_p$     & $E_{\mbox{\tiny min}}$  & background 
& $\sigma_{\mbox{\tiny stat}}$ & $\sigma_{\mbox{\tiny syst}}$ \\ 
& $[10^{32}]$ & [MeV] &   events  in 30 s & [MeV] & [MeV] \\
  \hline\hline
     &   &  &   & & \\[-1ex]
 Kamiokande-II & 1.4 & 7.5  & 0.55   & 1.27 & 1.0 \\
   `` &  `` & 4.5 & 5.6  & `` & `` \\
    IMB & 4.6 & 15 &  $0.01$  & 3.0 & 0.4 \\
  Baksan & 0.2 & 10 &  $1.0$  & 0.0 & 2.0 \\ \hline
 \end{tabular}}
 \caption{\em\small Summary of the main information regarding the detectors that is needed for data analysis. The  
 energy threshold of 4.5 MeV in Kamiokande-II allows  
 to analyze all 16 events in Tab.~\ref{tab0}.
See text for discussion. The background in IMB is 
 practically irrelevant. The error in Baksan is only described by the systematic component.
 \label{tab1}}
 \end{table}

Now we discuss certain expressions that considerably simplify  numerical analysis. 
The one regarding the total cross section has been derived
in \cite{viss} (see Eq.~24 therein) for energies that are relevant for a supernova.  It can be written as
 \begin{equation}\label{xsec}
 \sigma_{\mbox{\tiny IBD}}(E_\nu)=\kappa_{\mbox{\tiny IBD}}\ p_e E_e
  \mbox{ with } 
  {E_e=E_\nu-\Delta}
 \end{equation}
where the coefficient $\kappa_{\mbox{\tiny IBD}}$ depends mildly on $E_\nu$ as follows
\begin{equation}\label{kapp}
\kappa_{\mbox{\tiny IBD}}=e^{-0.07056\, x + 0.02018\, x^2 - 0.001953\, x^4} \times
 10^{-43}\mbox{cm}^2 ,\mbox{ where }x=\log[E_\nu/\mbox{MeV}]
\end{equation}
We use this parameterization to estimate the integral in Eq.~\ref{csec} by evaluating the integrand in the middle point
of its narrow range,
\begin{equation}
E_e^{\mbox{\tiny mid}}=\frac{E_\nu-\Delta}{1+E_\nu/m_p}
\end{equation} 
that we obtained by setting $\cos\theta=0$  in Eq.~\ref{midpoint}. 
When we let  
$d\sigma_{\mbox{\tiny IBD}}/dE_e=\sigma_{\mbox{\tiny IBD}}(E_\nu)\times \delta(E_e-E_e^{\mbox{\tiny mid}})$, the integral can be immediately solved, finding
 \begin{equation}\label{megl}
 \frac{dS_e}{dE_e}\approx \left. N_p
 \frac{dF}{dE_\nu}(E_\nu) \ 
 \sigma_{\mbox{\tiny IBD}}(E_\nu)\ J(E_\nu)\right. \mbox{ with }{E_\nu=\frac{E_e+\Delta}{1-\frac{E_e}{m_p}}} 
 \end{equation}
This is a rather accurate formula, that we will adopt 
in the numerical analysis.
Note that the last term of the previous equation is the Jacobian factor 
\begin{equation}
J=\frac{1}{| dE_e^{\mbox{\tiny mid}}/dE_\nu |}=
\frac{\left( 1+\frac{E_\nu}{m_p}\right)^2}{1+\frac{\Delta}{m_p}}
\end{equation}
that ensures that the total number of events--namely the 
integral  of 
Eq.~\ref{csec} over all possible $E_e$--is correctly reproduced.
 
 Finally,  we provide a simple numerical description of the angular distribution. Denoting by 
  $c$ the cosine of the angle between the incoming antineutrino and outcoming positron, we have
\begin{equation}\label{angolana}
\frac{1}{\sigma_{\mbox{\tiny IBD}}} \frac{d\sigma_{\mbox{\tiny IBD}}}{dc}\approx 
\frac{1}{2}\left[1+ 3\, c\, I_1+ \frac{5}{4} \left(3 c^2-1\right)\, (3 I_2-1) \right]
\end{equation}
where
$I_1=-0.0359+0.2391 x +0.0199 x^2$ and 
$I_2=0.3334 -0.0042 x +0.0258 x^2$, with $x=E_\nu/100\mbox{ MeV}$, that is again based on the calculation of \cite{viss}
and agrees with the expression  reported there, 
better than 1\% for all $5\mbox{ MeV}<E_\nu<50$ MeV.
It is actually considerably easier to implement for numerical calculations.


 \begin{figure}[t]
\centerline{\includegraphics[width=7cm]{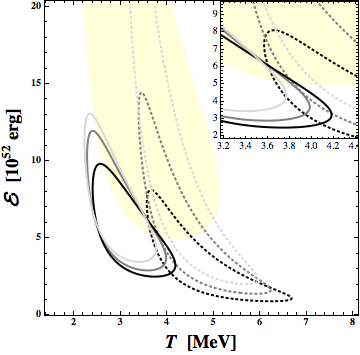}}
  \caption{\em\small Allowed regions for parameters, given 
  at 68.3\% C.L., obtained from analyses of the data of Kamiokande-II (3 continuous lines), IMB (3 dashed lines for $E_{\mbox{\tiny IMB}}=13,15,17$ MeV) and Baksan (shaded area). See the text for a detailed explanation on the various lines.
  \label{comp}}
 \end{figure}

 \section{Reference analysis \label{ra}}

 In this section, we describe our reference analysis of 
 the data from SN1987A. First of all, we discuss the compatibility of Kamiokande-II's, IMB's and Baksan's observations, Sect.~\ref{q1}; in the same section, the ambiguities of the analysis regarding the descriptions of the detectors are identified. 
Then, we present the combined analysis in Sect.~\ref{q2}, showing that the outcome is stable under several possible variants of that analysis, concerning the issues identified in Sect.~\ref{q1}. Finally, we discuss the implications of the best fit point in Sect.~\ref{q3} to provide the reader with an assessment and a few checks of our results.

 \subsection{Compatibility of the observations\label{q1}}
\begin{figure}[t]
  \centerline{\includegraphics[width=7.2cm]{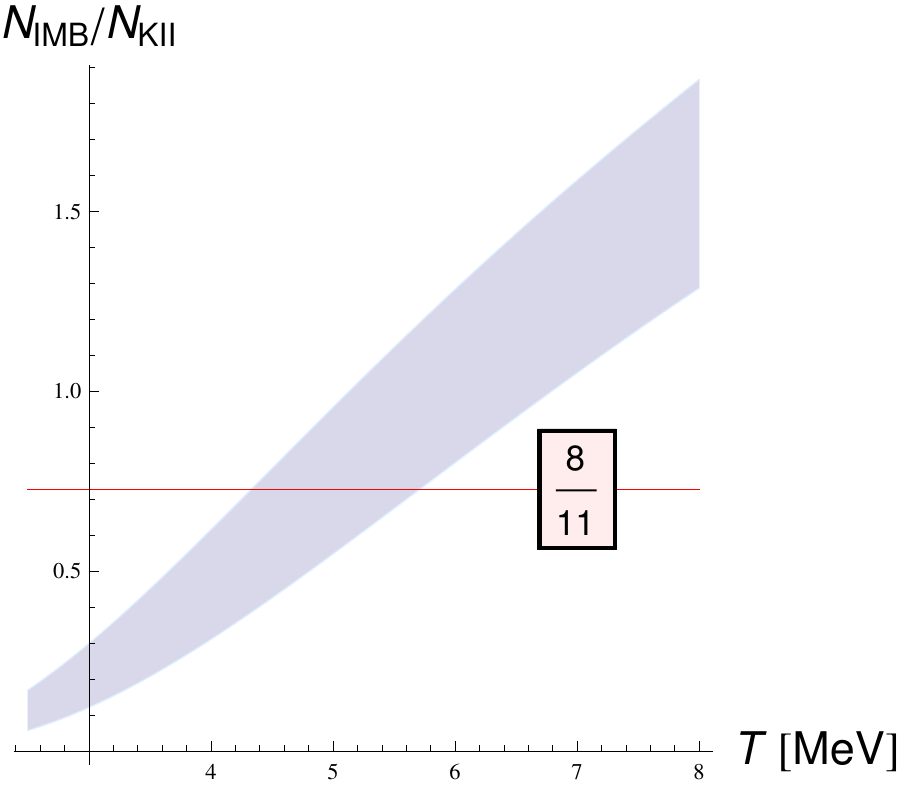} }
  \caption{\em\small  
  Comparison of the observed ratio of presumed signal events in IMB and Kamiokande-II, i.e. 8/11,  
  with the  expected value by varying  $E_{\mbox{\tiny \rm IMB}}$, given in Eq.~\ref{lgoggi}, in the range of Eq.~\ref{lpilo}.
  By diminishing  $E_{\mbox{\tiny \rm IMB}}$,  the concordance with the observed value  of the ratio obtains for lower values of $T$, that reproduces better the average energy of the Kamiokande-II dataset. 
  \label{ratunc}}
 \end{figure}

 Fig.~\ref{comp} displays various allowed regions, at 68.3\% C.L., obtained from the analysis of the single experiments. 
 More precisely,\\ 
1) The three continuous lines concern  Kamiokande-II; the uppermost line corresponds to the analysis of the 11 events with visible energy $E\ge E_{\mbox{\tiny min}}=7.5$ MeV and omitting the background, the next one is the same but includes the background for a time window of 15 seconds, the lower one is the analysis of the whole sample of 16 events in a time window of 30 seconds, see Tab.~\ref{tab0}. As we can see the three regions are compatible with each other; the inclusion of the background suggests that some (fraction of the) low energy events are not due to signal, thus shifting 
the temperature $T$ to somewhat higher values.\\
2) The three dotted lines concern IMB, and we go from the uppermost to the lowermost simply by decreasing $E_{\mbox{\tiny IMB}}$, see Eq.~\ref{lpilo}. This effect is more important than the previous one,  and can be explained qualitatively as follows. 
Decreasing $E_{\mbox{\tiny IMB}}$, the value of the efficiency increases. This increases the  number of events expected in IMB,  lowering the temperatures that are indicated by the observed event numbers ratio $N_{\mbox{\tiny IMB}}/N_{\mbox{\tiny KII}}$. In other words, decreasing $E_{\mbox{\tiny IMB}}$, the temperatures get closer to those indicated by the analysis of the data of Kamiokande-II. 
 This  is illustrated in Fig.~\ref{ratunc}.\\
3) The shaded area of Fig.~\ref{comp} corresponds to Baksan. Although this region is very wide, due to the rather limited amount of data, 
we see that for the 
lowest values of $\mathcal{E}$, 
the suggested values for $T$ 
are compatible and lie within those suggested by the other two experiments, as first pointed out by~\cite{ll}.\\

As it is clearly shown in the inset of Fig.~\ref{comp} the 68.3\% C.L.\ curves are quite close among them; in other words, the data are compatible with the same interpretation for some value of the parameters. This improves when we account for the possible occurrence  of background events in Kamiokande-II and if $E_{\mbox{\tiny IMB}}$ is decreased; moreover, the data of Baksan are not incompatible with those of the other two detectors, but rather improve the overall agreement.

 \subsection{Combined analysis of the data\label{q2}}
 
In view of the previous considerations,
we will consider from here on the  combined data set and use it to extract information on the parameters of electron antineutrino fluence. Tab.~\ref{tab2} gives the best-fit values for several combined analyses, that can be considered legitimate; it conveys the message that the  result of the statistical analysis is quite stable, which is reassuring.
 
 First of all, we note that 
there is no good reason to omit Baksan from the combined analysis, or use a narrow time and energy window for analysis in Kamiokande-II dataset. Then, we discuss in more detail the  apparently  `major' effect,  namely, the one due to systematic uncertainty in the efficiency of IMB. 

\begin{figure}[t]
\centerline{\includegraphics[width=6cm]{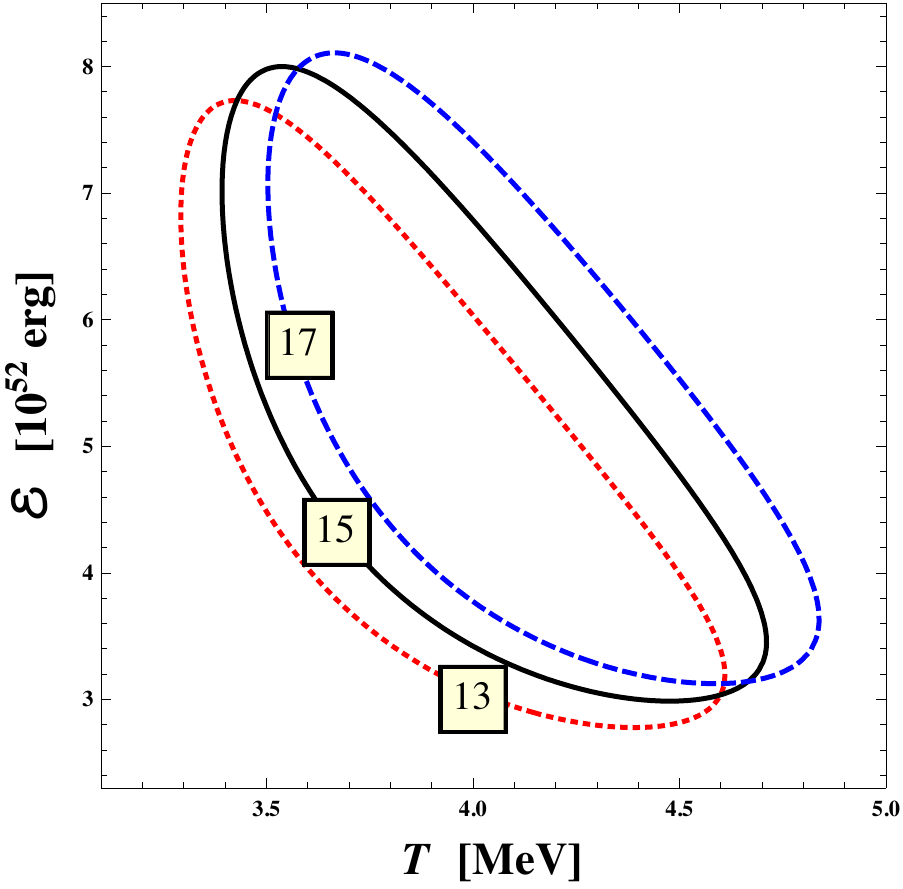}}
    \caption{\em\small 
 Allowed regions of the parameters from a combined analysis of all the data at the CL of 68.3\%, varying the efficiency of IMB within its range of uncertainty.  
 The values of $E_{\mbox{\tiny \rm IMB}}$ in MeV, as defined in Eqs.~\ref{lgoggi} and \ref{lpilo}, are indicated in the curves.
  \label{3cca}}
 \end{figure}

This is  illustrated in  Fig.~\ref{3cca}, that shows how the allowed parameter space changes with $E_{\mbox{\tiny IMB}}$: as we can see from Fig.~\ref{3cca}, the impact on the combined  dataset is not dramatic. Another way to illustrate this conclusion is to  consider the maximum value of the total likelihood $\mathcal{L}_{\mbox{\tiny max}}$: If  
we modify the reference efficiency of IMB,  for which $E_{\mbox{\tiny IMB}}=15$ MeV (Eqs.~\ref{lgoggi} and \ref{lpilo}), we find that 
$\mathcal{L}_{\mbox{\tiny max}}$ 
increases only by a factor 1.3 when we pass to $E_{\mbox{\tiny IMB}}=13$ MeV, while it decreases  by a factor 0.5 when we pass to $E_{\mbox{\tiny IMB}}=17$ MeV.

   \begin{table}[t]
\begin{center}
  \begin{tabular}{|ccc|cc|c|}
 \hline
 &&&& & \\[-1ex]
 $E_{\mbox{\tiny min}}(\mbox{KII})$ & $E_{\mbox{\tiny IMB}}$  & Baksan & $T$  & $\mathcal{E}$  & Comment on \\{}
   [MeV] &  [MeV] & analyzed  &  [MeV] & [$10^{52}$ erg] & 
   the analysis\\[1ex] \hline
 4.5 & 15 & yes & 4.0 & 5.0 & this paper \\ 
 4.5 & 15 & no & 4.2 & 4.1 & impact of Baksan \\
 4.5 & 13 & yes & 3.9 & 4.7 & higher IMB effic.\\
 4.5 & 17 & yes & 4.1 & 5.1 & lower IMB effic.\\ 
 7.5 & 15 & yes & 3.9 & 5.4 & other thresh.\ of KII \\ 
 7.5 (no bkgr.) & 15 & yes & 3.8 & 5.8 & other thresh.\ of KII \\
 7.5 (no bkgr.)& 15 & no & 3.9 & 5.0 & traditional \\ \hline
 \end{tabular}
 \end{center}
\caption{\em\small In the first 3 columns, we indicate which 
datasets are included and which type of analysis is performed. The resulting best fit values are given in the 4th and 5th columns. The last column is left for general remarks.  \label{tab2}}
 \end{table}

 For these reasons, we select as a {\bf reference analysis} the one where we use 1) a wide analysis region for Kamiokande-II  (energy threshold of $E_{\mbox{\tiny min}}=4.5$ MeV and  time window of 30 seconds); 2) the usual value of the efficiency in IMB, that corresponds to the choice 
 $E_{\mbox{\tiny IMB}}=15$ MeV, 3) the data of Baksan. For all the datasets we include a description of the background; this has a negligible role in the case of IMB.

 The best fit point,
 obtained analyzing the 29 events observed by the three detectors
 of Tab.~\ref{tab0}, is   
 \begin{equation}\label{piup}
 \mathcal{E}=5\times 10^{52}\mbox{ erg}\ ,\ T=4\mbox{ MeV}
 \end{equation}
  Curiously,  the reference values agree almost perfectly with  the outcome of the traditional and simpler analysis based on 19 events: see Tab.~\ref{tab2}. This point will be illustrated further in the next section.

 
The total likelihood function depends upon the two  parameters that summarize the astrophysics of the emission,  
$\mathcal{E}$ and $T$. 
  Subsequently, we can obtain a pair of functions of a single parameter by integrating away the other one. These functions are the  
one-dimensional likelihoods, thanks to the fact that the total  likelihood is non-negative and it gives 1 when integrated over all possible values of the parameters. From the one-dimensional likelihoods, we can derive the   regions that are compatible with the analysis of SN1987A observations (i.e.\ the allowed ranges) at any desired confidence levels ($C.L.$). For instance, at 68.3\% $C.L.$ 
  (i.e.\ at 1$\sigma$) we have
\begin{equation}
\mathcal{E}=4.8^{+2.3}_{-1.0}\times 10^{52}\mbox{ erg , }
T=3.9^{+0.5}_{-0.3}\mbox{ MeV}
\end{equation}
this shows that the errors we estimate from SN1987A's data analysis are not very large.

 The  allowed regions for the two parameters 
 can be obtained by imposing a
Gaussian condition corresponding to 2 degrees of freedom.
This condition can simply be written as\footnote{A Gaussian distribution with 2 d.o.f.\ 
$ \mathcal{L}(x_1,x_2)=G(x_1,\sigma_1) G(x_2,\sigma_2)$ has
$
\int  \mathcal{L}(x_1,x_2) dx_1 dx_2=
\int_0^{u_*}  e^{-u} du=1- e^{-u_*}<C.L.$, that implies  
 $1-C.L.<e^{-u_*}= \mathcal{L}(x_1,x_2)/ \mathcal{L}{\mbox{\tiny max}} $, i.e.\ Eq.~\ref{gausian}.}
\begin{equation}\mathcal{L}(\mathcal{E},T)>\mathcal{L}_{\mbox{\tiny max}} \times (1-C.L.),
\label{gausian}\end{equation} 
and the regions resulting from the application of this procedure  are adopted all throughout this work.

 \begin{figure}[t]
\centerline{\includegraphics[width=7cm]{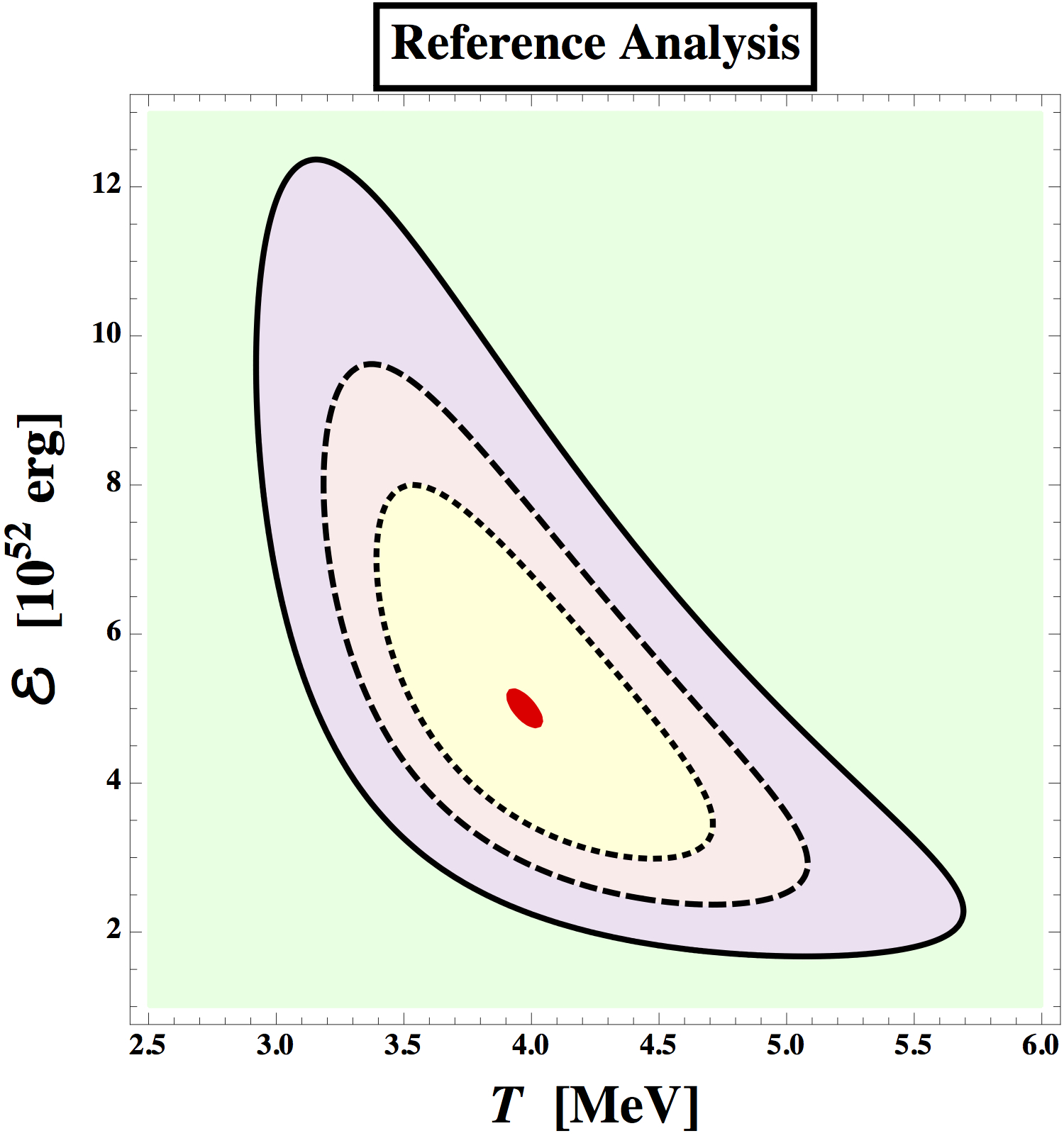}}
    \caption{\em\small {\bf Reference analysis:}
  Allowed regions of the parameters from a combined analysis of the data at the C.L.\ of 1\%, 68.3\%, 90\%, 99\%.
  \label{comb}}
 \end{figure}

This statistical procedure is quite adequate; e.g.\ when we integrate the likelihood of our reference analysis in the region defined by Eq.~\ref{gausian} setting
$C.L=0.5$,  we find 49.6\% of the total integral, that differs from 50\% by only 0.7\%; when we set $C.L=0.95$,  the fraction gets even closer, with a difference smaller than 0.1\%. For our purposes, such a small difference is negligible and we prefer to adopt the  Gaussian procedure, that is significantly more convenient from the computational point of view.

The allowed regions for the parameters from the reference analysis are given in Fig.~\ref{comb}.
  We set $E_{\mbox{\tiny IMB}}=15$ MeV in Eq.~\ref{lgoggi} and lower the threshold for Kamiokande-II's analysis to 
  $E_{\mbox{\tiny min}}(\mbox{KII})=4.5$ MeV; thus,
  all events of 
  Tab.~\ref{tab0} 
  are included in the reference analysis.

\subsection{Characteristics of the best fit point\label{q3}}
Here we would like to check {\em a posteriori} the properties of the best fit solution when it is compared with the individual data sets.

 The expected number of signal events in the best fit point is 
 \begin{equation}\label{nombo}
 N_{\mbox{\tiny KII}}=14.4, \ 
 N_{\mbox{\tiny IMB}}=5.9, \ 
 N_{\mbox{\tiny BAK}}=1.8
 \end{equation}
 for a total of 22.1 signal events, compatible with
 the number $21.4\pm 1.2$
 derived in Sect.~4 of \cite{vulcano}. 
 
 When we sum the numbers of Eq.~\ref{nombo}
 and the expected numbers of background 
 events, given in Tab.~\ref{tab1}, we can evaluate the Poisson probability $P(\ge N)$ to observe at least the number of events $N$ of Tab.~\ref{tab0}. We find
 \begin{equation}P_{\mbox{\tiny KII}}=84\%,\  
 P_{\mbox{\tiny IMB}}=24\%,\ 
 P_{\mbox{\tiny BAK}}=16\%
 \end{equation}
 that illustrates the plausibility of the best fit point.
 Note that the presence of background events in Baksan (Tabs.~\ref{tab0} and \ref{tab1}) is crucial 
 to account for the 5 observed events without invoking a statistical fluctuation with $P< 5$\%.

 We have also checked the plausibility of the resulting spectra using the  Smirnov-Cram\'er-von Mises (SCVM) test,
 \begin{equation}
 \omega^2=\frac{1}{12\ N_{\mbox{\tiny obs}}}+\sum_{i=1}^{N_{\mbox{\tiny obs}}} \left[F(E_i) -\frac{i-1/2}{N_{\mbox{\tiny obs}}}\right]^2
  \end{equation}
  where $F$ is the cumulative distribution of events 
  in function of the visible energy including signal and background.
  In the best fit point of the reference analysis we find that 
  \begin{equation}
 \alpha_{\mbox{\tiny KII}}=31\%,\  
 \alpha_{\mbox{\tiny IMB}}=22\%,\ 
 \alpha_{\mbox{\tiny BAK}}=58\%
  \end{equation}
  where we have used the formulae for $\alpha(\omega^2)$ given in \cite{ad} with the corrections for a finite set of events given in \cite{st}.
Again, the test on the spectrum ensures that the outcome of the fit is perfectly acceptable. 

\subsection{Gamma lines?\label{q4}}

As a final test of the reference analysis, 
we would like to quantify the expected   
gamma ray signals from 
neutral-current neutrino interactions, that lead to 
excited nuclei. 
 We will argue that, although it is not possible to exclude that this happened,  the chances  are not very high at the relevant neutrino energies.  For the temperature of the electronic antineutrinos 
 we 
 adopt  the best fit value of Eq.~\ref{piup} $ T=4$ MeV, and 
 vary the common temperature of all non-electronic neutrino and antineutrino species within $T_{x}=(1-1.3)\times T$ \cite{keil,carol}.

 First, we discuss the excitation of ${}^{12}$C leading to
 a monochromatic line at  $E_\gamma=15.11$~MeV.
In fact,   the dataset of Baksan includes three events 
(B1, B2 and B4 of Tab.~\ref{tab0})
 that are compatible with this interpretation.
According to \cite{carol}, the number of expected events at the best fit point is rather small, $S=(0.05-0.12)$. Noting that the value of the uncertainty function  (Eq.~\ref{erri} and Tab.~\ref{tab1}) is $\sigma=3$ MeV, the  chance that no event of Baksan was due to neutral current excitation of carbon is estimated 
to be $\mathcal{P}=(80-90)~\%$. 

Second, we consider the two lines at $\sim 5$ and $6$ MeV that result from excited  ${}^{15}$O and  ${}^{15}$N, produced 
by stripping off a nucleon from ${}^{16}$O \cite{lei}. This is potentially interesting for the interpretation of some of the low energy events of  Kamiokande-II, especially, the five  events K6, K13, K14, K15, K16 of Tab.~\ref{tab0}. Adopting the approximate cross section $0.9 \times10^{-44} \times [(E_\nu/10\mbox{ MeV})-1]^5$ cm$^2$ summed over neutrino and antineutrinos \cite{lei,cei}, and the average
value $\eta\sim 0.7$, 
we find a signal of 
$S=(0.03-0.07)$ events for each line. Using  $\sigma=1.5$ MeV for both lines, the  chance that none of the above 5 events was due to neutral current excitation of oxygen  is estimated to be $\mathcal{P}=(90-95)~\%$. 

Therefore, it is not possible to exclude that one observed event was due to these reactions,  but the hypothesis adopted for the reference analysis (i.e.\ only IBD)
is not  contradicted and remains quite plausible. 
Note incidentally that for the purpose of data analysis 
it is more important to account for the presence of background events, than of these reactions.

 \section{Alternative analyses\label{aa}}

The data of SN1987A have been studied and analyzed since they appeared. The first detailed statistical analyses were published  several months after the events. Two reference works regarding the study of fluence  are those published in 1988 by  Bahcall et al.~\cite{bahcall} and in 1996 by Jegerlehner et al.~\cite{jeger}. In this section, we examine these and other procedures of analysis, and in Sect.~\ref{cc}  we will compare their outcomes  with the one of the reference analysis, Sect.~\ref{ra}.

 We recall the various traditional procedures adopted to analyze these data, Sect.~\ref{p1};  we consider several different efficiency uses in Sect.~\ref{p2}; we show how to include the 
 remaining experimental information in the analysis of SN1987A observations in Sect.~\ref{p4}; finally, we discuss  two 
 modified  forms of the electron antineutrino fluence in Sect.~\ref{p3}.

 \subsection{Traditional procedures\label{p1}}
 In this section, we summarize several traditional procedures that are usually implemented in the analysis of SN1987A observations. Namely: data selection, approximated description of the signal, errors in measurements processing.
 \subsubsection{Selection of the dataset \label{p11}}
 The traditional analyses, as \cite{bahcallBook} or \cite{jeger},
 use a selected part of the dataset in order to confidently exclude the possible occurrence of background events.\footnote{This approach has been discussed critically in the literature \cite{loredo1}, arguing that it is not possible to confidently attribute any individual event to the signal, and  strictly speaking the choice of region of analysis (volume, energy, time) has been made  {\em a posteriori}. These considerations make it appealing to treat the data as conservatively as possible and suggest the relevance of an accurate description of the signal.}
 The data that are used in the traditional analyses are:
 1) those of IMB; 
2) those of Kamiokande-II, collected in the entire inner volume of the detector in a time window of less than 15 seconds and 
 above the threshold used for solar neutrino analyses, $E_{\mbox{\tiny min}}=7.5$ MeV.
  This makes 8+11=19 events in total. 
The data of Baksan, subject to background contamination, are simply discarded. 


 \subsubsection{Signal description\label{p12}}

 The traditional processing can be obtained beginning from  Eq.~\ref{midpoint},
  setting 
  $\delta\approx \Delta=m_n-m_p$ 
  (that is an excellent approximation) and 
  considering the limit $m_p\to \infty$. In this approximation, 
  the fluence is calculated at the point $E_\nu=E_e+\Delta$ and can be extracted from the integral
  in Eq.~\ref{csec}. In this way, we obtain the expression
 \begin{equation}\label{tred}
 \frac{dS_e}{dE_e}\approx \left. N_p
 \frac{dF}{dE_\nu}(E_\nu) \ 
 \sigma_{\mbox{\tiny IBD}}(E_\nu)\right. \mbox{ with }{E_\nu=E_e+\Delta} 
 \end{equation}
 that coincides with Eq.~\ref{megl}
 when the formal limit $m_p\to \infty$ is considered.  
 Another element of the traditional data processing is the use of the inverse beta decay cross section in the form of Eq.~\ref{xsec}
 but using a constant value of $\kappa_{\mbox{\tiny IBD}}$. E.g.~the value  adopted 
 in  \cite{jeger} (without references to the previous literature) is 
 $\kappa_{\mbox{\tiny IBD}}=2.295\times 10^{-44}\times \mbox{cm}^2/m_e^2$.
  A more precise but still simple expression of the cross section can be obtained using Eq.~\ref{kapp}, namely letting $\kappa_{\mbox{\tiny IBD}}$ to be a function of $E_\nu$. Moreover, the description of $dS_d/dE_e$ can be improved further using Eq.~\ref{megl}, implying the inclusion of terms order $E_\nu/m_p$ (as discussed in Sect.~\ref{sibd}) that play some role for high energy neutrinos.

Using the case where 
$T=4$ MeV as an example, we checked that with Eq.~\ref{kapp} the total number of events is reproduced within 0.1\%.  The  
approximation of Eq.~\ref{tred} 
exceeds the correct one 
by $-3$\%, $-1$\%, +7\% and +21\% at $E_e=10$, 20, 30 and 40 MeV respectively, while the new treatment of  Eq.~\ref{megl} adopted in this work exceeds it only by +0.6\%, +0.9\%, +0.4\% and $-1.4$\%. That is a considerable improvement.

\subsubsection{Individual errors\label{p13}}

In principle the error, the intrinsic efficiency and the background rate
do not only depend on the energy, but also, e.g.\ on the position of the events in the detectors.  This kind of considerations explains the residual discrepancy of Fig.~\ref{sigm}. Thus, one may wish to avoid the use of analytical description of the uncertainty functions $\sigma(E_e)$, and  
take into account the individual nature of the error for each single event. (Incidentally, we are not aware of whether it is possible to perform something similar for intrinsic efficiency and background processing by correcting the values event by event;
see \cite{jcap} for further discussion.)


 \subsection{Other ways to use the efficiency\label{p2}} 
In order to determine 
the likelihood function of Eq.~\ref{lugu} fully, we 
need to specify two quantities. They are:
 1)~the total number of events above 
a certain visible energies $E>E_{\mbox{\tiny min}}$, i.e.\ $N_{\mbox{\tiny tot}}$;
1)~the number of signal events for a certain value of  the visible energy $E_i$,  i.e.\ $dN_i$.
 In the formalism we described in Sects.~\ref{purea} and \ref{forma},   
 the first quantity contains the total efficiency,
 whereas 
 the second one contains the intrinsic efficiency  function. Other studies use the (total and intrinsic) efficiency in different ways; according to the present formalism, this should be considered as  bias. 
 
We compare our use of the efficiency with the one adopted in the two existing analyses of the flux \cite{ll,pagl}. 
In Sects.~\ref{p21} and \ref{p22}, we outline how their treatment of the efficiency departs from the one that is
adopted in this paper, described in Sects.~\ref{purea} and \ref{forma}.
 
 \subsubsection{Total number of signal events\label{p21}}
In order to calculate the total number of events, Eq.~\ref{tuttic},
it is possible to use the published efficiencies directly. However, this 
requires to use the same energy threshold that has been adopted for the analysis by the experimental collaboration; this is done e.g.\ in \cite{jeger}.
Suppose that we need to lower the energy threshold, in order to separate signal from background 
 {\em a posteriori} rather than {\em a priori}. In this case, the total efficiency should be changed, as discussed in Sects.~\ref{purea} and \ref{forma}. 
However, such a correction has not been done in \cite{ll,pagl}. 
This is relevant for the analysis of Kamiokande-II data; note that this  experimental collaboration mentions explicitly the fact that the total efficiency changes  by changing the analysis threshold~\cite{kam}.  

 \subsubsection{Differential number of signal events\label{p22}}
The second point regards $dS/dE$, see Eq.~\ref{petella}. 
Two different prescriptions have been adopted: in \cite{ll} where the intrinsic efficiency $\eta$ is replaced by 1, and in \cite{pagl} where $\eta$ is replaced by the total efficiency $\epsilon$ taken from the publications of the experimental collaborations. 
From the standpoint of the formalism described in Sect.~\ref{forma}, the right answer stays in the middle. Thus, one should conclude that 
these alternative prescriptions introduce another bias concerning  
the analysis of IMB data in \cite{ll} (since the intrinsic efficiency $\eta$ deviates strongly from 1),   and a bias  concerning  
the analysis of Kamiokande-II data of \cite{ll} (since the intrinsic efficiency $\eta$ does not coincide with the total one).
%

 \subsection{Impact of measured angle and time on the analysis\label{p4}}
 
In this section, we discuss how we should include the experimental information on the time and direction of the events in the analysis of the fluence.
 
\subsubsection{Time distribution (flux) \label{p41}}
In principle, one could analyze the flux, i.e. time differential distribution, of neutrinos, rather than the fluence. 
This has been done in a couple of analyses, see \cite{ll,pagl}, finding in both cases a 2-3 sigma hint for an increased initial luminosity,
consistent with the expectations on the  emission during the initial phase of rapid accretion onto the nascent neutron star. 
Here, we will simply multiply Eq.~\ref{flenza} by a function
$\ell(t)$ normalized to 1 when integrated between $t=0$ and $t=30$ s and at the same time we divide the background rate by 30 s. More precisely we will considered the following two cases
\begin{equation}
 \ell(t)\propto \left\{
 \begin{array}{lr}
\exp[-t/3.5\mbox{ s}]   & \mbox{exponential cooling} \\
8 \exp[-t/0.4\mbox{ s}] +  [1+t/(4.5\mbox{ s}) ]^{-1.5} & \mbox{model with accretion component}
 \end{array}
 \right.
\end{equation}
In both cases, signal events happening in earlier times have a higher chance of being due to signal rather than background.

\subsubsection{Angular distribution\label{p42}}

The following steps have been done in order to account for each event's scattering angle:
1)~we estimated {antineutrino energy} from positron energy $E_e$ and the cosine of the scattering angle  $c=\cos\theta$, 
using Eq.~\ref{midpoint}; 
2)~we replaced the total cross section $\sigma_{\mbox{\tiny IBD}}(E_e)$ with $ d\sigma_{\mbox{\tiny IBD}}(E_e,c)/dc$ in Eq.~\ref{megl}, 
and at the same time halved the background;
3)~we neglected the experimental error on  $c$ (checking that it was adequate);
4)~we included the intrinsic angular efficiency (=bias) for IMB, 
$1+0.1 c$ \cite{imb}. These assumptions simplify numerical calculations, and one can somewhat decrease the burden by using the approximate distribution of Eq.~\ref{angolana}.
In principle some of the events could be due to other reactions, but our cursory investigations of this hypothesis did not offer sufficient motivations to justify a special analysis. 

Before continuing, 
we would like to explain the reasons for the last statement,  
for the sake of the reader interested in this specific issue.
Consider the Kamiokande-II 
dataset with visible energy $E_{\mbox{\tiny min}}\ge 7.5$ MeV,
that is dominated by IBD events;
the {\em a priori} probability  to have one  
elastic scattering event is  $\sim 30$\% \cite{prdML,astro}.
A single elastic scattering event would improve the agreement of the angular distribution with the expectations. E.g.~according to~\cite{prdML}, 
the SCVM goodness of fit value passes from 8.6\% to 26.7\%, that at first sight motivates further discussion. The candidate elastic scattering event is the first one, K1 of Tab.~\ref{tab0}, but this is not emitted exactly in the direction of the neutrinos  and has a rather high energy. 
Using the fluences of \cite{keil}, its chance to be   due to 
elastic scattering is found to be  $\sim 30$\%  \cite{astro}. 
This number remains about the same when we compare $\bar\nu_e$ with $\langle E_{\bar\nu_e}\rangle=12$ MeV and 
$\bar\nu_e$ with $\langle E_{\nu_e}\rangle> 20$ MeV (a speculative value, that would maximize the agreement with K1), if 
$\nu_e$ and $\bar\nu_e$ have the same luminosities. Instead, the chance  diminishes with neutrino oscillations, since
the $\nu_e$ are converted into 
$\nu_{\mu,\tau}$ which have weaker
interactions.


 \subsection{Modified form of the fluence\label{p3}}
 Here, we discuss two physical reasons 
 to modify the fluence  of Eq.~\ref{flenza}
 adopted elsewhere in this work,
 and quantify that plausible amount of modification, before studying its impact on the analysis of the data. These physical effects are the `pinching' of the emitted spectra and the flavor transformations due to neutrino oscillations.
 
 These cases cover a range of possibilities which is wide enough to illustrate the impact of a modified flux on the analysis of SN1987A data. 
 However, it should be emphasized that for certain applications (e.g.\ the search for the neutrinos emitted during the entire lifetime of the Universe) some specific features of the spectrum, such as the high energy tail of the distribution, can particularly be important. 
 In other words, we have to judge case by case whether the assumed shape of  the electron antineutrino distribution is adequate for the application in which we are interested.

\subsubsection{Pinching\label{p31}}
Various analytical arguments and analyses of the numerical simulations suggest that the flux of all types of neutrinos, electron antineutrinos included, is not entirely thermally distributed. In particular, it has been found that the high energy tail is less populated. This has been described analytically by means of various parameterizations of quasi-thermal distributions. 
 The most recent and apparently most flexible parameterization of the flux $\Phi=dF/dt$ is \cite{keil}
 \begin{equation}
 \frac{d \Phi}{dE_\nu} =  \frac{L}{4\pi D^2} \times 
 \frac{E_\nu^\alpha\ e^{-E_\nu/{T}_\star}}{\Gamma(\alpha+2)\  T_\star^{\alpha+2}} 
 \mbox{ where }
 T_\star=\frac{3}{\alpha+1} \times T
 \end{equation}
 where $L$ is the luminosity (i.e., the radiated power), so that $\int_0^\infty L(t) dt=\mathcal{E}$.
  The meaning of ${\mathcal E}$ is the same as before, and
  the quantity $T_\star$ allows us to maintain the same expression 
  for   the average energy of electron antineutrinos $\langle E_\nu \rangle=3 T$, Eq.~\ref{smpq}, that we use 
  all throughout this paper.
  The width of the distribution, defined as
  \begin{equation} 
  \delta E_\nu/\langle E_\nu \rangle=\sqrt{ 
  \langle E_\nu^2 \rangle/\langle E_\nu \rangle^2-1 }
  \end{equation} 
  is given by the simple expression
  \begin{equation}\frac{{\delta E_\nu}}{{\langle E_\nu \rangle}}=\frac{1}{\sqrt{1+\alpha}}\end{equation}
  and decreases by increasing the parameter $\alpha$, or in other words, the spectrum becomes more `pinched' than in the 
case   chosen as reference, Eq.~\ref{flenza}, for which
$\alpha=2$  and thus 
 $\delta E_\nu/\langle E_\nu \rangle=1/\sqrt{3}$.

The typical value quoted for antineutrinos is
 $\alpha\approx 3$ \cite{keil};  but when we integrate over the time, in order to pass from flux $\Phi$ to fluence $F$, we have to take into account that all astrophysical quantities, including luminosity and temperature, change with time. Thus, the fluence is a convolution of quasi-thermal distributions which can still be approximated by a quasi-thermal distribution but with a value of $\alpha$
 closer to the reference one. In short, the fluence is `more thermal' than the flux. Interestingly, this remark was made on general theoretical ground in \cite{jeger},  and recently it was  put on quantitative bases by a detailed fit of the flux using SN1987A data \cite{aanda}.  

\begin{figure}[t]
\centerline{
\includegraphics[width=7cm]{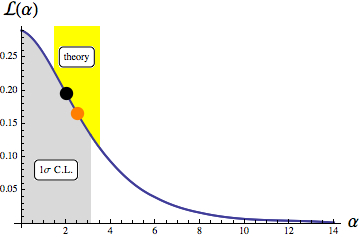}
\includegraphics[width=7cm]{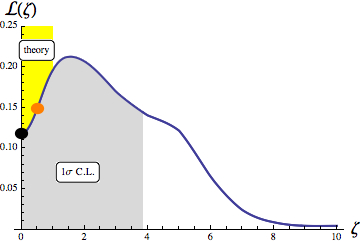}}
  \caption{\em\small Left (resp., right) plot: Likelihood as a function of the pinching parameter $\alpha$ (resp., oscillation parameter $\zeta$), varied in a range much larger than suggested by the theoretical expectations.
The $1\sigma$ allowed ranges of parameters  $\alpha<3.1$ 
(resp., $\zeta<3.8$), grey regions, 
overlap with the conservative expectations
$1.5\le \alpha \le 3.5$ (resp.,  $0\le \zeta\le 1$),
yellow bands.
The  circles show the values assumed here:
black for the reference analysis of Sect.~\ref{q2}, orange for the 
cases of comparison discussed in Sect.~\ref{p3}.
  \label{pipolin}}
 \end{figure}

Some analyses of the SN1987A data have speculated on pinching factors that depart a lot from  the  expectations  \cite{bcg,mr}. 
Admitting for the sake of the argument  
all possible positive values of 
$\alpha$ in the analysis, 
we have checked that the  
best fit value of the likelihood 
decreases mildly with $\alpha$ but  the preference for $\alpha<2$ is not significant and the only sound 
statistical inference is $\alpha<3.1$ at 68.3\% C.L.:
See Fig.~\ref{pipolin}.

In view of the above considerations, we assume the following form of the fluence 
  \begin{equation}
 \frac{dF}{dE_\nu} =  \frac{\mathcal{E}}{4\pi D^2} \times 
 \frac{E_\nu^\alpha\ e^{-E_\nu/{T}_\star}}{\Gamma(\alpha+2)\  T_\star^{\alpha+2}} 
 \mbox{ where }
 T_\star=\frac{3}{\alpha+1} \times T
 \end{equation}
using the case $\alpha=2.5$ in the analysis, that approximates well the results of~\cite{aanda} and does not deviate strongly  
from theoretical expectations.

Let us conclude with a 
theoretical consideration. At the best of present theoretical knowledge, one believes that $\alpha$ is slightly larger than 2.
At the same time, we have discussed the non-significant indication 
for smaller values of $\alpha$, and also the value suggested by the analysis of the flux 
which is not very far from 2. 
In view of these considerations, and also the discussion on the role of oscillations in the next 
section, 
we think that the assumption $\alpha=2$, adopted in the traditional \cite{bahcall,jeger} and in our reference analysis of SN1987A data should be considered rather reasonable. 
  
 \subsubsection{Oscillations\label{p32}}
The effect of flavor conversion, usually called `oscillations', is to mix the emitted spectra, that we will identify with the superscript ${}^0$ for clarity. For instance, electron antineutrino fluence on Earth is 
\begin{equation}\label{oscass}
\frac{dF}{dE_\nu}=  P\times \frac{dF^0_{\bar\nu_e}}{dE_\nu}+ (1- P)\times \frac{dF^0_{\bar\nu_x}}{dE_\nu} \mbox{ with }
P=0.67
\end{equation}
where $x$ denotes a non-electronic neutrino and where $P=P_{\bar\nu_e\to \bar\nu_e}=|U_{e1}^2|$ is 
the survival probability.\footnote{ 
In the conventional decomposition,  
$U_{e1}=\cos\theta_{12}\cos\theta_{13}$. 
A global analysis of the data gives 
$\cos^2\theta_{12}=0.692\pm 0.017$ and $\cos^2\theta_{13}=0.976\pm 0.002$ \cite{capozzi}, thus 
 $P=0.676\pm 0.017$, whose  error is dominated by $\cos^2\theta_{12}$.}
The very simple formula for 
$P$ applies to the case of normal mass hierarchy and also to the case of inverted hierarchy, when we use the description of collective 
flavor transformations of \cite{mdd}. If instead the collective flavor transformation are inhibited, 
the case of inverted hierarchy leads to $P\to 1$ \cite{sd}, that coincides with the case of no oscillations (only the interpretation of the parameters of Eq.~\ref{flenza} changes: in this case, they are the parameters of non-electronic neutrinos). 
See \cite{md} for more discussion of oscillations.

We have neglected the Earth matter effect in the analysis, since its impact is quite small~\cite{cavanna}.  
Also, note that 
the effect of collective flavor transformations could be quite complex, leading e.g.\  to spectral swaps above or below some threshold; in principle, we could have time-dependent flavor transformations; etc. 
While these effects are very interesting theoretically, it is not clear to us whether the present understanding of this phenomenon can be considered  conclusive. 
Our main purpose is just to show that, when we adopt  
Eq.~\ref{oscass} and a constant value of $P$, consistent with the experimental indications, the impact of flavor transformations on the analysis of the electron antineutrino fluence from 
SN1987A is not large. 
Moreover, it is plausible that similar effects will increase the average value of $P$ over the energy and the time, relevant for the analysis of the fluence, thereby decreasing the impact of flavor conversion on the  electron antineutrino signal. 
(See \cite{first,pagl} for a discussion of the potential impact on the first second of neutrino emission.)

In the case when the fluences $F^0_{\bar\nu_e}$ and $F^0_{\bar\nu_x}$ differ only for the amount of radiated energy the situation is trivial 
since we should simply set $\mathcal{E}\equiv P \mathcal{E}_{\bar\nu_e}+ (1-P) \mathcal{E}_{\bar\nu_x}$.
Instead, in the case when these energies are similar or equal (equipartition) but the temperatures are different we will set
\begin{equation}
T_{\nu_x}=(1+\zeta)\times T_{\bar\nu_e} \mbox{ and }
T_{\bar\nu_e}=  \left[ 1-   (1-P) \frac{\zeta}{1+\zeta}\right]\times \frac{3 }{\alpha+1}\times T
\end{equation}
where the first relationship quantifies the difference 
of the temperatures (i.e., it is the definition of $\zeta$), and the second one allows us 
to maintain  the identification $\langle E_\nu\rangle=3 T$,  
Eq.~\ref{smpq}, as everywhere else in this paper. 

It is quite intuitive that, by broadening the distribution of neutrinos, the oscillations produce an `anti-pinching', or with the notation of the previous paragraph, they will diminish the effective pinching of the spectrum. This can also be understood from the following analytical formula,
\begin{equation}\label{letto}
\frac{{\delta E_\nu}}{{\langle E_\nu \rangle}}=\frac{1}{\sqrt{1+\alpha}}\times \sqrt{1 + (2+\alpha) P (1-P)\times \frac{\zeta^2}{1+\zeta} }\end{equation}
In the past, $\zeta$ was believed to be quite large; this parameter was considered completely unconstrained in some analysis of the data \cite{lnrd}, finding hints of even larger values of $\zeta$, strongly deviating from the values suggested by the conventional astrophysical picture. In our calculations, we find that this hint is not statistically significant, see
Fig.~\ref{pipolin} and
compare with the discussion of the previous section.

Moreover, the most recent calculations \cite{macheccazz} suggest that $\zeta$ is close to zero. 
In order to illustrate--and plausibly maximize--the impact of the oscillations on data analysis without 
strongly contradicting 
the expectations, we will use a relatively large value $\zeta=0.5$ together with $\alpha=2$. 
Note that, if we have $\zeta=0.5$ 
and  $\alpha=2.5$ and use Eq.~\ref{letto} we find  
 $\delta E_\nu/\langle E_\nu \rangle\approx 1/\sqrt{3}$ which is the value for the reference case 
that does not include either pinching or oscillations.
This remark implies the two effects, if  
included together, would lead to a partial compensation. 

Note in passing that we focus on the discussion of oscillations of 
electron antineutrinos, namely, of the main species involved in the interpretation of the events observed by Kamiokande-II, IMB and Baksan. For other neutrino species such as electron {\em neutrinos} which could be important in alternative scenarios (including some speculative  interpretations of the events of LSD) or presumably for the interpretation of future observations, the formulae are different 
and their quantitative impact can be much more relevant. See e.g. \cite{sd,pocc}.

   \section{Comparison of the allowed regions\label{cc}}
\begin{figure}[t]
\centerline{\includegraphics[width=5.cm]{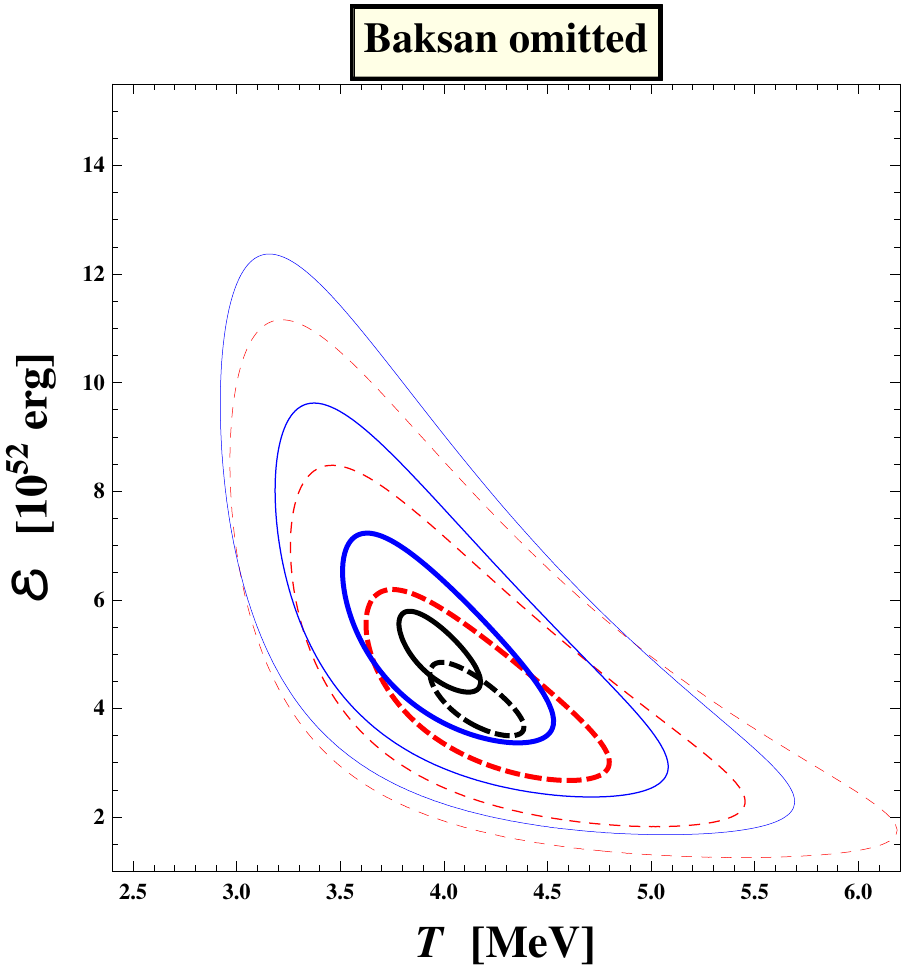}\includegraphics[width=5.cm]{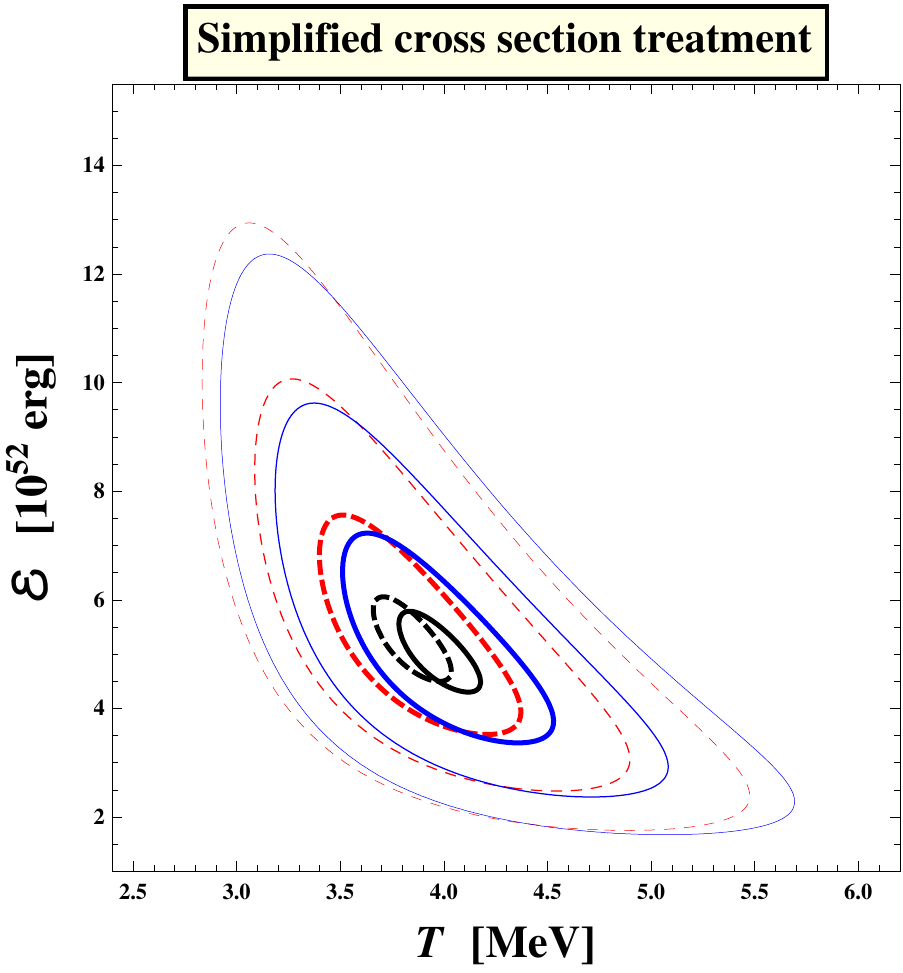}\includegraphics[width=5.cm]{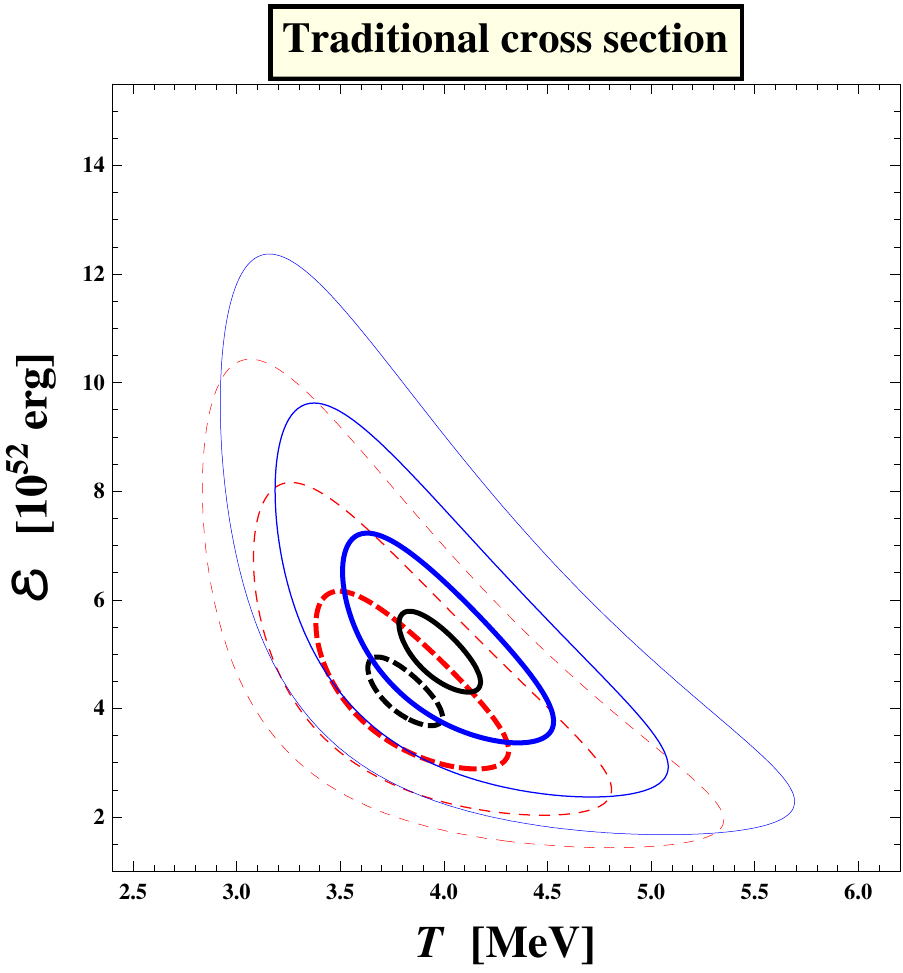}
}
\vskip4mm
\centerline{\includegraphics[width=5cm]{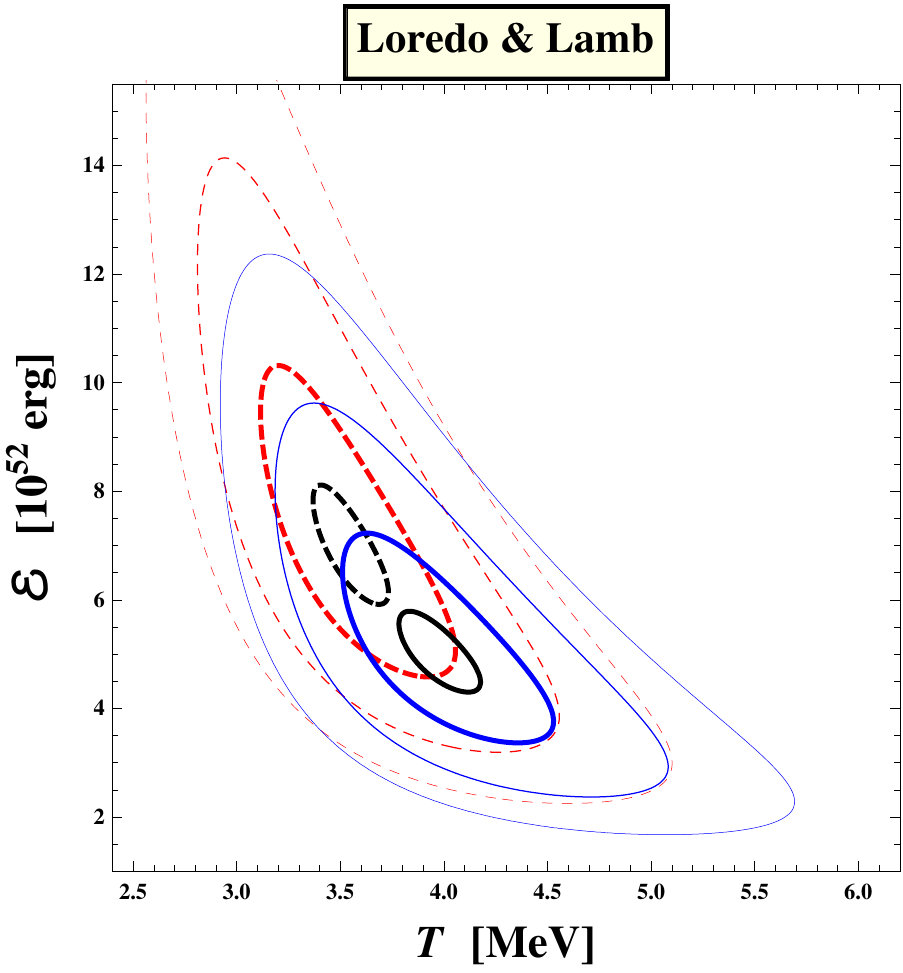}\includegraphics[width=5cm]{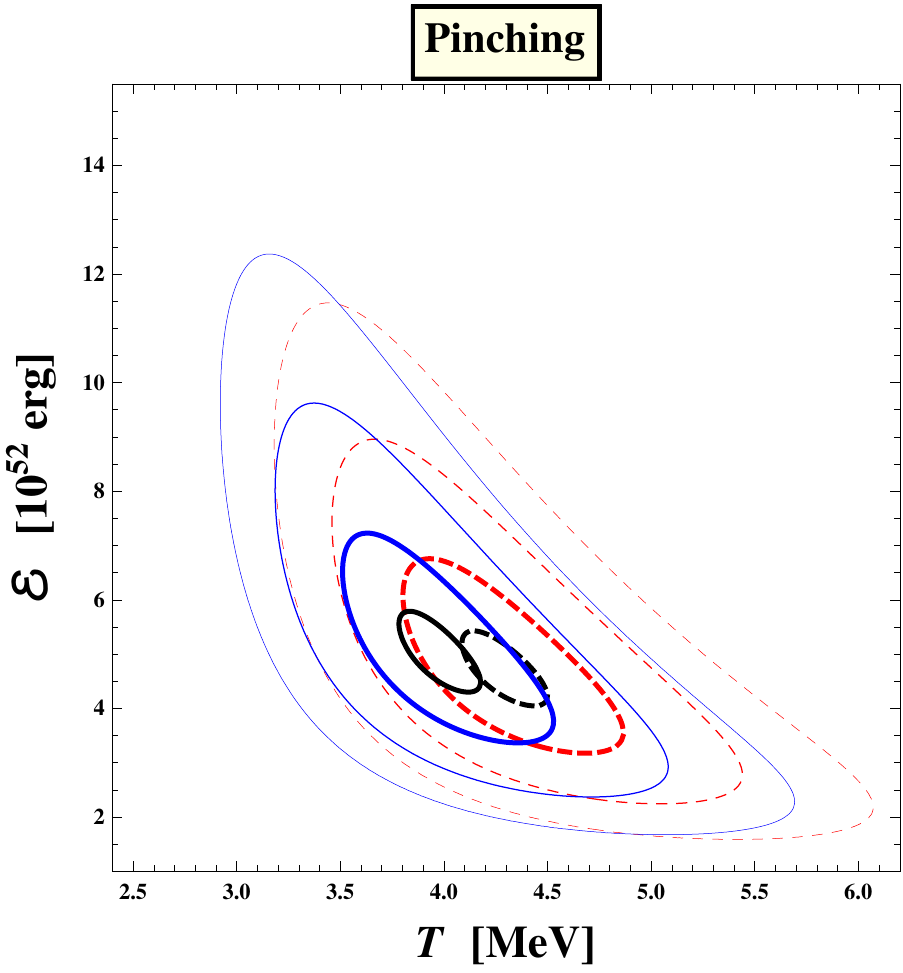}\includegraphics[width=5cm]{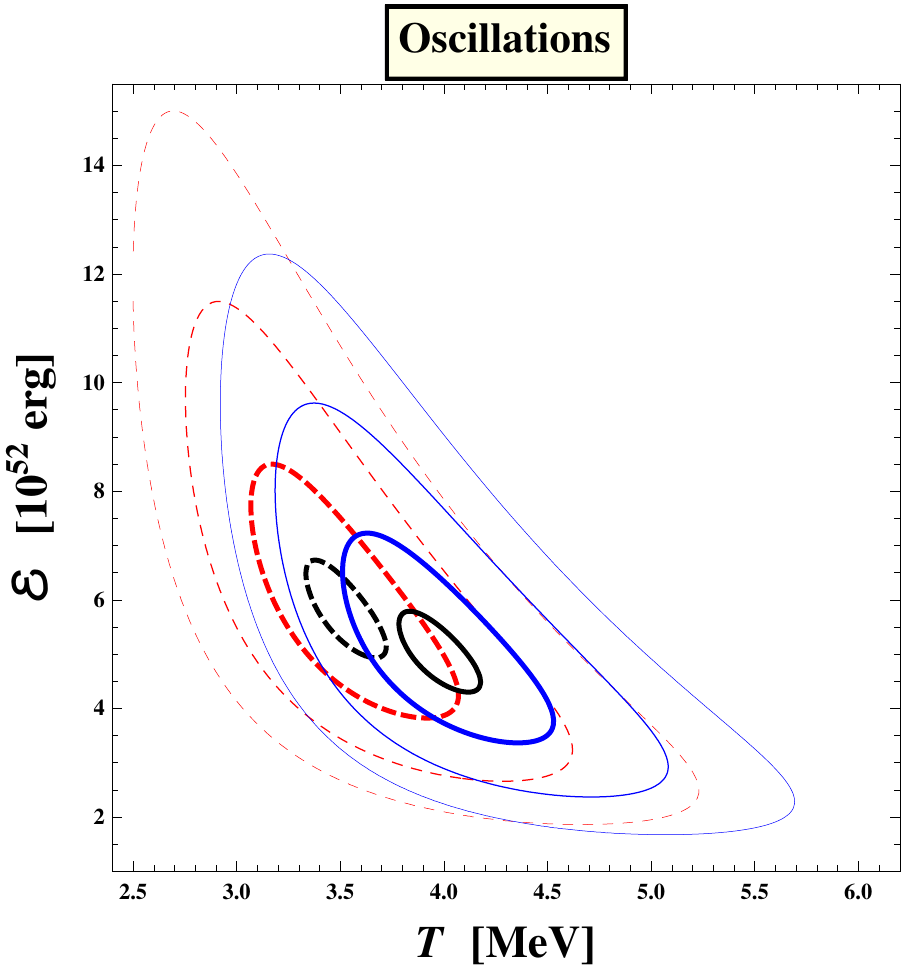}}
\caption{\em\small Comparison of the reference analysis of Sect.~\ref{ra}
(continuous lines) with six variations (dotted lines) that imply  perceivable shifts  of the allowed regions. See the text of Sect.~\ref{secb} for a detailed description.\label{figb}}
\end{figure}

\begin{figure}[t]
\centerline{\includegraphics[width=5cm]{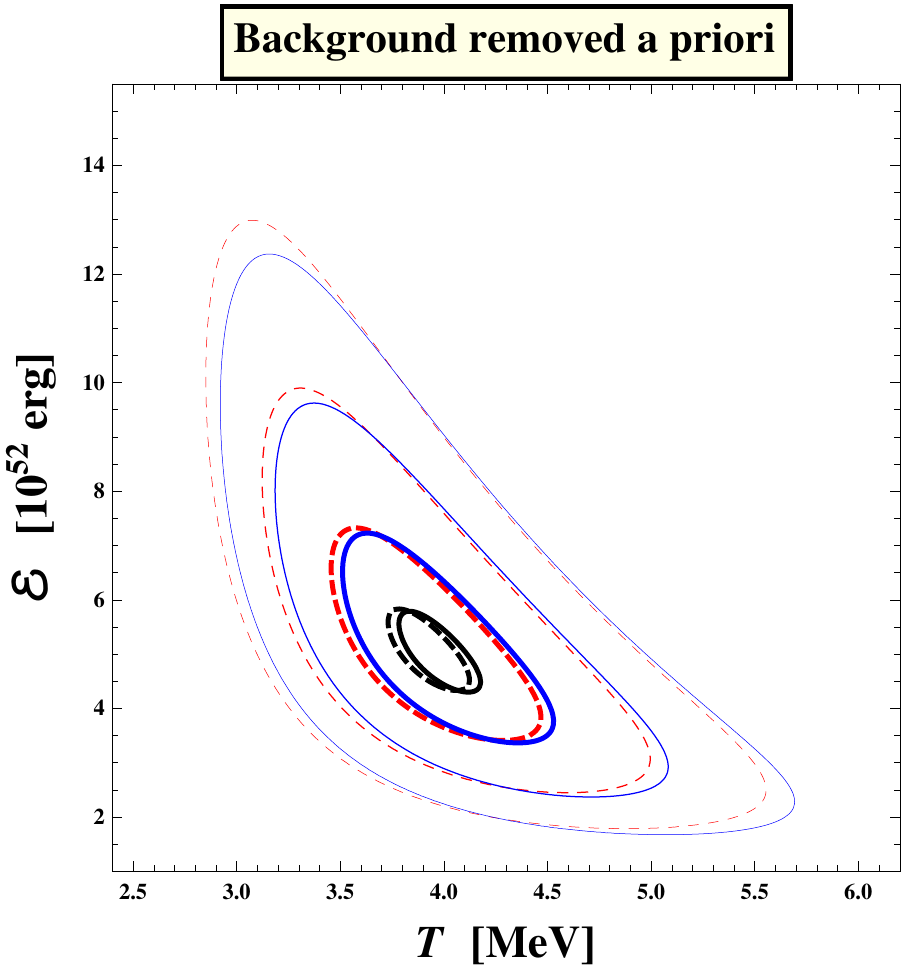}\includegraphics[width=5cm]{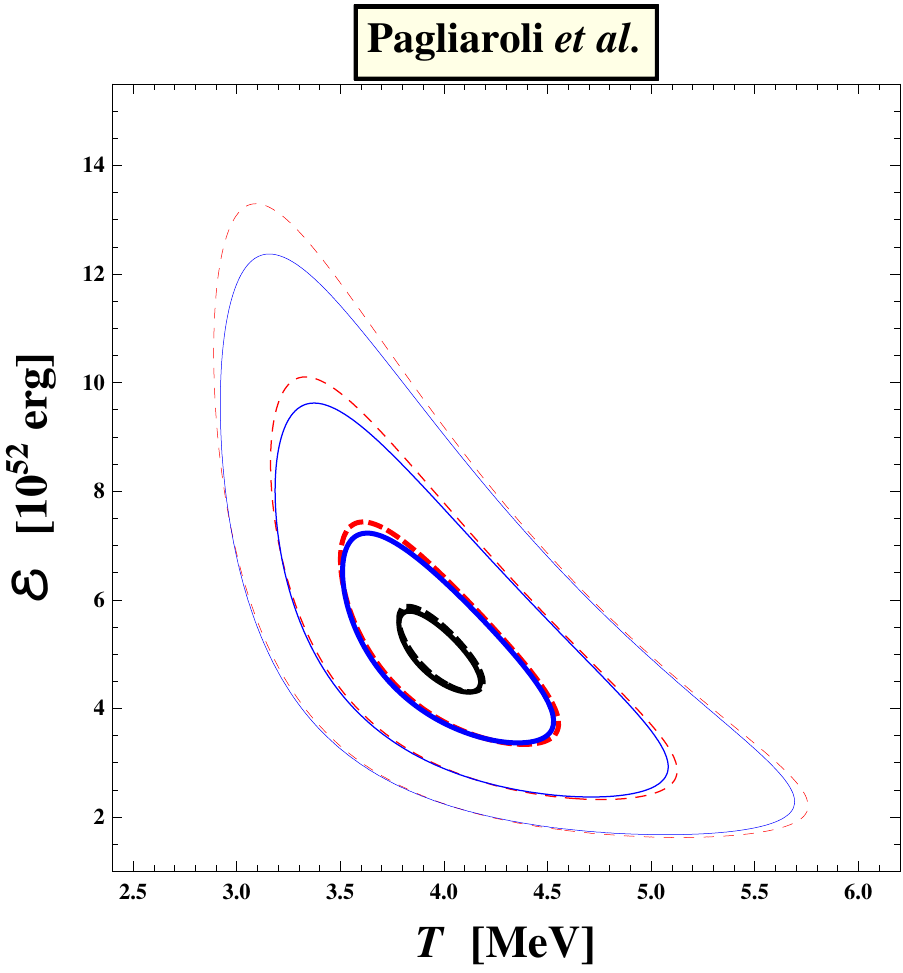}\includegraphics[width=5cm]{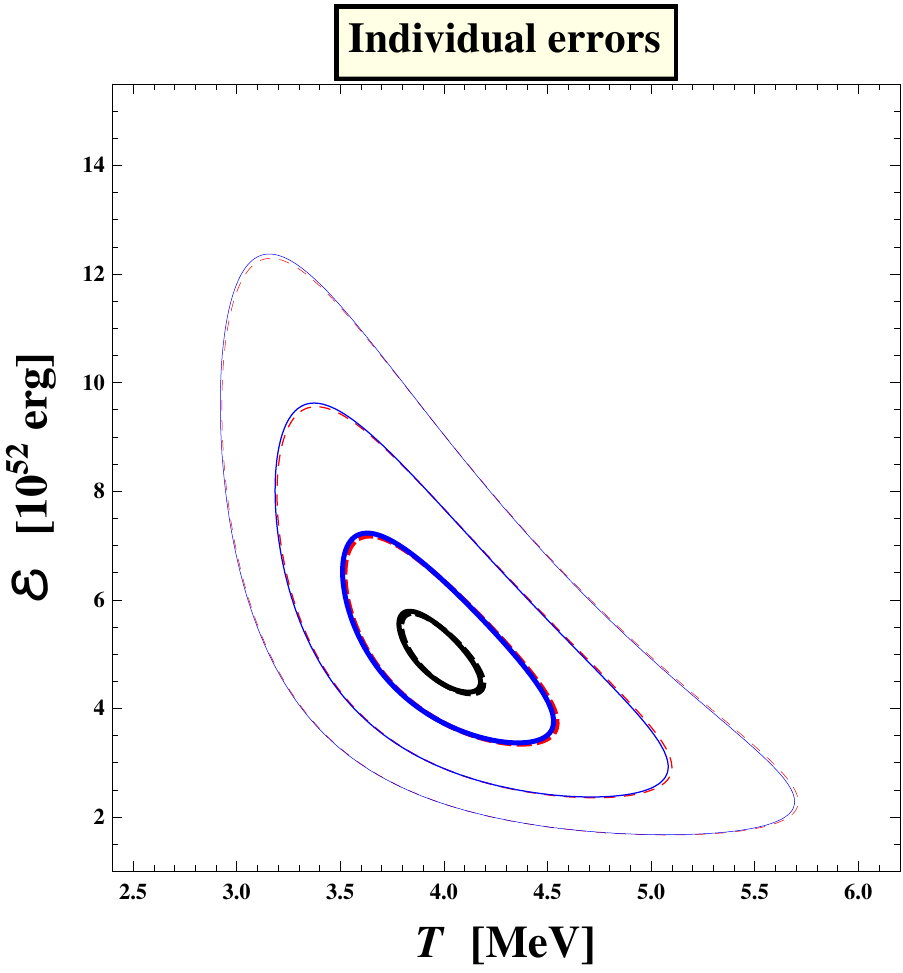}}
\vskip4mm
\centerline{\includegraphics[width=5.cm]{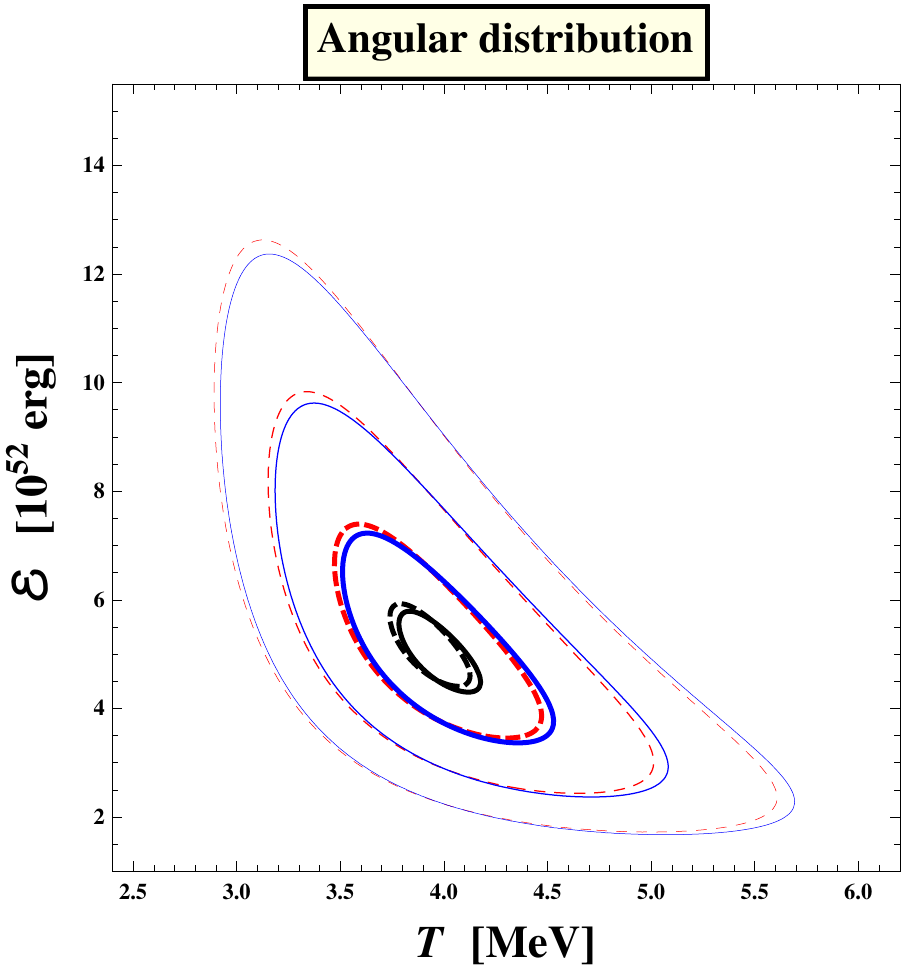}\includegraphics[width=5.cm]{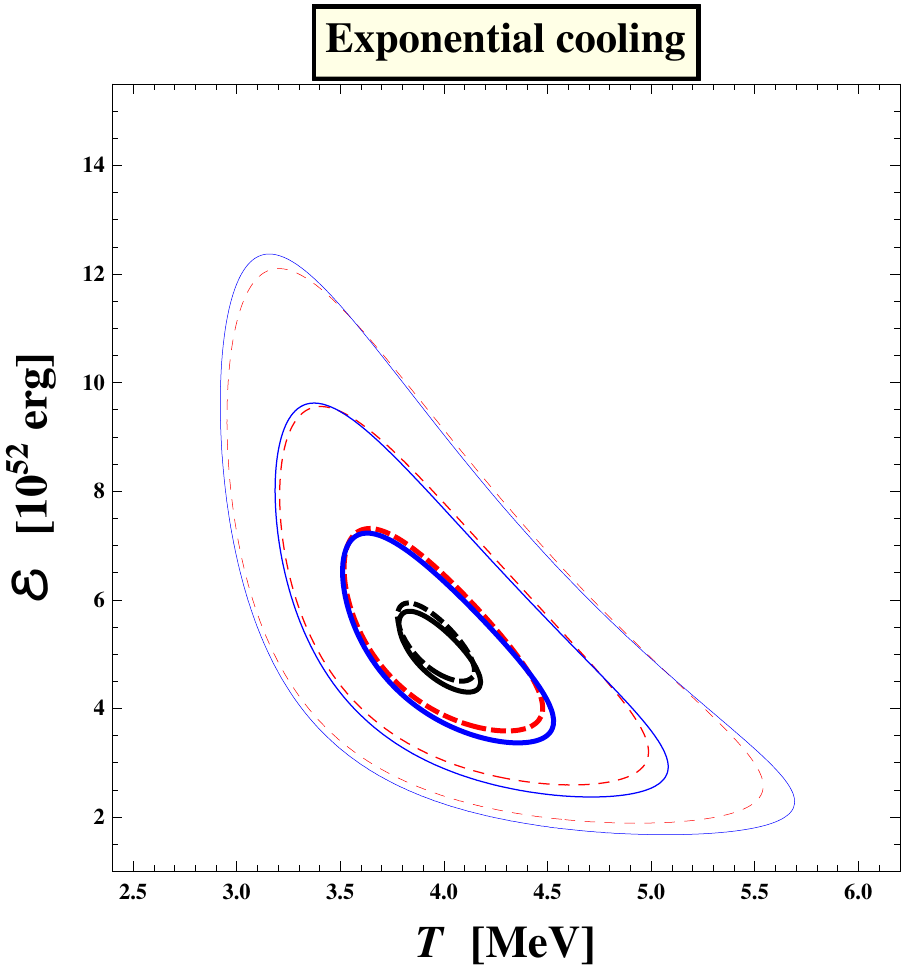}\includegraphics[width=5.cm]{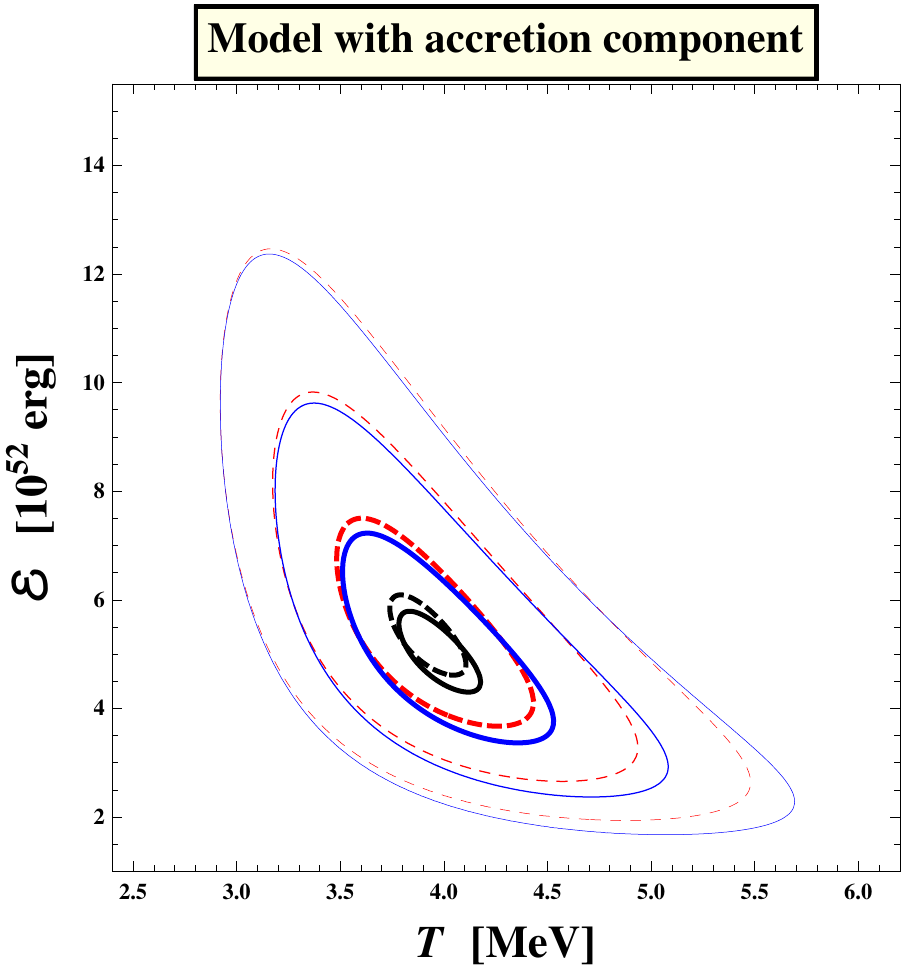}}
\caption{\em\small Comparison of the reference analysis 
 of Sect.~\ref{ra} (continuous lines) with six variations (dotted lines) that imply only minor modifications of the allowed regions though. See the text of Sect.~\ref{secc} for a detailed description. \label{figc}}
\end{figure}

 After the discussion of all possible modifications, we are ready to quantify the effect. We do it in this section, by comparing the allowed regions for the parameters $T$ (the temperature and/or 
 1/3 of the average neutrino energy)
 and $\mathcal{E}$ (the energy carried by the electron antineutrinos) see Eq.~\ref{flenza}. We show the regions at the following confidence levels
  \begin{quote}
  10\%, 50\%, 90\%, 99\%
 \end{quote}
 as estimated with a Gaussian procedure, Eq.~\ref{gausian}. We compare the 
 results of Sect.~\ref{q2} (continuous lines) and those of alternative analyses in Figs.~\ref{figb} and \ref{figc} (dashed lines). There are 12 cases, corresponding to the plots of Figs.~\ref{figb} and \ref{figc}. While Fig.~\ref{figb} shows the modifications of some relevance, Fig.~\ref{figc} describes the cases where the relevance is small or negligible. Let us illustrate the comparison,  discussing each of these 12 cases separately and 
  following the order of presentation:

 \subsection{Perceptible shifts: Fig.~\ref{figb} \label{secb}}
 
 \paragraph{Baksan omitted} 
 This is discussed in Sect.~\ref{p11}; the relevance of these data have been argued in 
 \cite{ll}.  By omitting Baksan, one biases somewhat the analysis in favor of higher temperatures and lower radiated energies, as shown in 
the upper-left plot of Fig.~\ref{figb}.

  \paragraph{Simplified cross section treatment}
  This is discussed in Sect.~\ref{p12}: it is the treatment of the cross section as described in Eq.~\ref{tred} along with the precise expression of the total cross section of Eq.~\ref{kapp}. 
  The comparison is shown in the upper-middle plot of Fig.~\ref{figb}, and the effect is a mild bias in favor of lower temperatures.

    \paragraph{Traditional cross section}
  This is discussed in Sect.~\ref{p12},
   and used e.g.\ in \cite{bahcall,jeger}. It is the same as the previous point, but using the cross section of Eq.~\ref{xsec}, that is about 30\%  larger than the correct one.
  The comparison is  shown in the upper-right plot of Fig.~\ref{figb} 
  and the bias is a bit larger than the one in the previous point; note in particular the decrease of $\mathcal{E}$.
 
  \paragraph{Loredo \& Lamb}
  This  has been introduced in \cite{ll}, and
  is discussed in Sect.~\ref{p2} 
   and in ref.~\cite{fernando}.
  The modifications, shown in the lower-left plot of Fig.~\ref{figb}, 
   is of some importance, both in $T$ and in $\mathcal{E}$ and it is mostly due to the bias introduced in the analysis of IMB data.

   \paragraph{Pinching}
   This has been  first pointed out in 
   ref.~\cite{jh} and it is 
   discussed in Sect.~\ref{p31}. As we see from the lower-middle plot of Fig.~\ref{figb} in presence of a moderately pinched spectrum,
   $\alpha=2.5$, higher energies are somewhat  favored.

    \paragraph{Oscillations}
  Oscillations of SN1987A antineutrinos, introduced in ref.~\cite{alesha}, have been largely discussed since then. 
  The  last plot of Fig.~\ref{figb} shows an effect similar in size but opposite in direction to the previous one. Indeed,   
  as argued in Sect.~\ref{p32}, oscillations  counterbalance pinching, 
  especially with the selected parameters $\alpha=2.5$ and  $\zeta=0.5$.

  \subsection{Minor variations: Fig.~\ref{figc} \label{secc}}

  \paragraph{Background removed {\em a priori}} 
 This is discussed in Sect.~\ref{p1},  and consists in the 
 omission of Baksan and low energy Kamiokande-II events, and in
 barring {\em a priori} the possibility of having 
 background events; see e.g.\ \cite{bahcall,jeger}.
 As shown in the first (upper, leftmost) plot of Fig.~\ref{figc}, the region allowed for the astrophysical parameters are almost the same. This is due to a compensation between the effect of the inclusion of Baksan and the upgraded analysis of Kamiokande-II, as suggested  by Tab.~\ref{tab2}, by the left plot in Fig.~\ref{comp} and by  Fig.~\ref{figb}.
  
  \paragraph{Pagliaroli {\em et al.}}
 This is used in \cite{pagl} and it 
 is discussed in Sect.~\ref{p2}. 
The modification (upper-middle plot) is small due to a compensation of the two causes of bias discussed in Sects.~\ref{p21} and \ref{p22}. As a consequence, the numerical results of Ref.~\cite{fernando} are almost unaffected, despite the conceptual difference with the procedure outlined in Sect.~\ref{ra}.

  \paragraph{Individual errors}
  This is discussed in Sect.~\ref{p13} and mentioned in 
  \cite{jeger}. As  we see from the upper-right plot of Fig.~\ref{figc}, this is absolutely negligible in agreement with \cite{jeger}. In view of the discussion of Sect.~\ref{p13}, this is good news since we are not loosing anything essential using an `average' description of the response of the detector as in Sect.~\ref{forma}.

   \paragraph{Angular distribution}
    This is considered in Sect.~\ref{p42} and analyzed in 
  \cite{prdML,pagl}. Again, the effect shown in the lower-left plot of Fig.~\ref{figc} is negligible. This means that the puzzling aspects of the angular distribution (especially those of IMB \cite{prdML}) have no significant impact on the determination of the parameters.

    \paragraph{Exponential cooling}
    This is discussed in Sect.~\ref{p41} and mentioned in 
  \cite{ll,pagl,fernando}. The effect of a non-trivial time distribution on the region of the allowed parameters is minor, as we see from the lower-middle plot.

     \paragraph{Model with accretion component}
  This is discussed in Sect.~\ref{p41} and mentioned in 
  \cite{ll,pagl}. The lower-right plot and the previous one suggest that the study of the time distribution can be to some extent  decoupled from the determination of the temperature and of the total radiated energy.

  \section{Discussion\label{culdis}}

  Before concluding, we would like to 
  compare the results obtained from the fit with a simple description of the energy spectrum (Sect.~\ref{o2}); examine the 
 correspondence with the theory in Sect.~\ref{o3};
 mention the possibility of more complex 
 analyses of these data in Sect.~\ref{o35};
and finally  discuss 
 the major open questions (Sect.~\ref{o4}).

\subsection{Comparison with a 
`model independent' description\label{o2}}

In \cite{jcap}, the question of how to describe the signal spectrum maximizing the role of the observations and keeping model dependence to a minimum was considered. This discussion has been then developed  in \cite{bur}. 

\begin{figure}[t]
\centerline{\includegraphics[width=8cm]{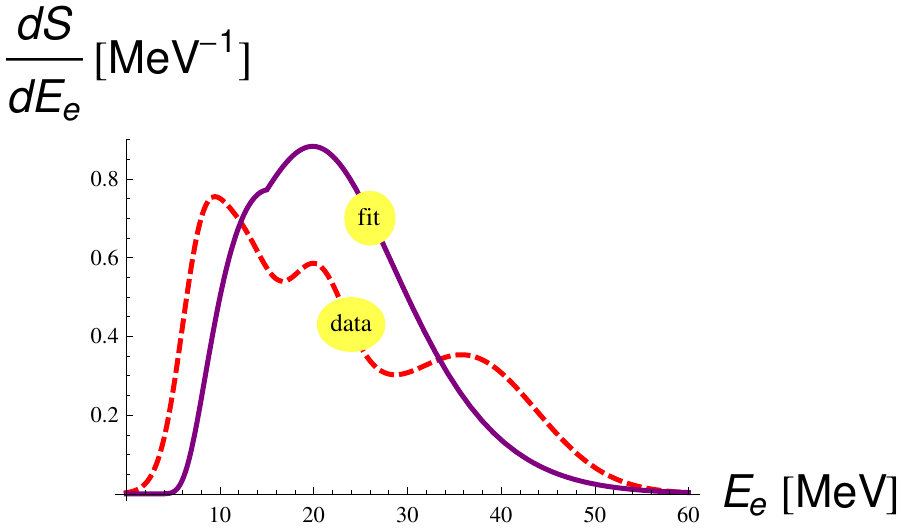}}
    \caption{\em\small Comparison of the curve obtained by the data of Kamiokande-II and IMB, Eq.~\ref{guess} (dashed line)
 with the spectrum resulting from the best fit fluence,  
 Eq.~\ref{giust} (continuous line).
  \label{lupoblu}}
\end{figure}

A simple analytical description has been proposed \cite{jcap,bur} as a way to present the data,
taking advantage of the observed energies $E_i$ and their errors $\sigma_i$. It is the following function of the true energy $E_e$,
\begin{equation}\label{guess}
\left. \frac{dS}{dE_e} \right|_{\mbox{\tiny data}}=\sum_{i=1}^{N_{\mbox{\tiny obs}}} G(E_e-E_i,\sigma_i)
\end{equation}
The corresponding 
curve is smooth enough to look reasonable, 
even though it tends to overestimate the role of fluctuations.\footnote{Indeed,  
such a construction does not lead to anything useful in the case of the angular distribution, that is known {\em a priori} to be wider;  
see also Sect.~\ref{o3}.} 
It is instructive to compare this curve with the expectation for the signal at the best fit point  obtained in this paper. This can simply be written as
\begin{equation}\label{giust}
\left. \frac{dS}{dE_e} \right|_{\mbox{\tiny fit}}=A_{\mbox{\tiny eff}}(E_e)\times \frac{dF}{dE_\nu}(E_\nu)
\end{equation}
where we took advantage of Eqs.~\ref{tuttic} and \ref{megl}
to introduce the  {\em effective area}
\begin{equation}
A_{\mbox{\tiny eff}}(E_e)=N_p\ \sigma_{\mbox{\tiny IBD}}(E_\nu)\, J(E_\nu)\,  \epsilon(E_e,
E_{\mbox{\tiny min}}) 
\end{equation}
and where $E_\nu$, as the function of $E_e$, is given in Eq.~\ref{megl}.

Using only the data of IMB and those of Kamiokande-II with $E_{\mbox{\tiny min}}\ge 7.5$ MeV in order to select 
{\em a priori} a relatively clean data set of {\em signal} events, and inserting  
the best fit value of Eq.~\ref{piup} in the fluence of Eq.~\ref{giust},
we  arrive at the curves that are shown in Fig.~\ref{lupoblu}.
The integrals agree rather well, since 
the first momentum is 19 for  Eq.~\ref{guess} and 19.7 for 
Eq.~\ref{giust}, and the average energies are 
22.3 MeV and 22.4 MeV in both cases respectively.

The  comparison of this figure emphasizes that the description based on the data
(i.e. on Eq.~\ref{guess}) has a few more events 
 than expected at low energy (Kamiokande-II) and at high energy (IMB), with a consequent depletion in the intermediate energy region.\footnote{This corresponds closely to the moderate tension visible in Fig.~\ref{comp} and the 
 discussion regarding the effect of 
 oscillations and (anti)pinching as well, see Sect.~\ref{p3}.} 
 However, these features are not remarkable 
 and considering the quantitative discussion given in this work, 
 the agreement between the two curves can be considered satisfactory. The summary of the data provided by the  
 fit is less prone to fluctuations and takes advantage of the expectations, thus in our view, it should be preferred. (Compare with discussion in Sect.~\ref{p3} as well).

\subsection{Comparison with the theory\label{o3}}

Now, we would like to compare the results that we have obtained 
with the theory. First of all, there are general issues. E.g.~it is legitimate to ask: was SN1987A a standard supernova? but also: what is a standard supernova? Or at least: what are the characteristics of  neutrino emission from a standard supernova? In our humble opinion,
such issues would deserve further discussion. We believe that it would be desirable to plan systematic theoretical investigations, keeping in mind the needs of neutrino observatories. 

Next, we come to the specific theoretical question: how do the results, in particular the best fit of Eq.~\ref{piup}, compare with  the general theoretical picture.  
If we estimate the luminosity of a neutrino black body emitter with temperature $T$ and radius $R$, we find 
$L= 3/\pi\times R^2 \times T^4$  
(the proportionality constant being exact 
for the distribution of Eq.~\ref{flenza} but not crucial for the argument). 
Using $T=4$ MeV and   
$R=15$ km, close the radius of a neutron star,
we find $L=3.4\times 10^{51}$ erg/s. Thus, using again the best fit result,  $\mathcal{E}=5\times 10^{52}$ erg, 
we deduce the emission time
$\tau=\mathcal{E}/L=15$~s, that is in the expected range: 
it compares well with the observations, it is not far from a typical diffusion time in the core. 

Even more simply, we can note the agreement of $\mathcal{E}$ with the general expectation on the radiated energy, 1 sixth of $\sim 3\times 10^{53}$ erg, and also the agreement of the temperature with the result of the 
 most recent calculations of the average energy of the antineutrinos \cite{jankocc}. Summarizing, the agreement of the expectations with the best fit point is impressive and  might even seem suspiciously too good. However this is a rather stable indication of the analysis, as discussed in Sect.~\ref{cc}.

It is instructive to recall that the present agreement has been reached only recently. We had persistent claims in the past 20 years, that the value of average antineutrino energy has to be about 15 MeV, thus significantly larger than the one indicated by SN1987A \cite{bahcallBook,truran} and that the discrepancy would significantly increase in presence of neutrino oscillations with large mixing angles \cite{alesha,jeger}. Apparently,  both claims were based on  predictions that turned out to be incorrect or in other words 
 were caused by an underestimation of theoretical errors. 

In this connection, it is interesting to note that we do not have a formal assessment of the theoretical errors yet, in particular, the errors on the predictions of $\mathcal{E}$ and $T$. 
This is obviously due to the theoretical difficulties of the problem, and can be contrasted with the situation of another type of 
low energy neutrinos that have been successfully detected, i.e.  
solar neutrinos: when these were first observed, the predictions with error bars of Bahcall were already available.
Perhaps, it will be possible to have reliable assessments of theoretical errors before the next galactic supernova. 

It is not clear what one should do in these conditions, but  caution is advisable. In particular, it seems questionable that one could blindly rely on the latest theoretical calculations of the fluxes and fluences, assuming the errors are negligible. In fact, making predictions adopting the fluxes and the fluences deduced from SN1987A in the meantime could be considered an alternative safer practice, since (some) errors can be estimated in this way. 
In this work, the theoretical input has been deliberately kept 
to a minimum, testing 
 the assumed form of the electron antineutrino fluence
 for adequacy {\em a posteriori}.

Conversely, the analysis supernova neutrinos is very relevant for nuclear astrophysics. Here are few  illustrations of this statement: (1)   neutrinos with a temperature of $\sim 4$ MeV allow to successfully produce certain elements observed in nature and otherwise unaccountable \cite{kajino}; (2) the observation of  electron antineutrinos offers us a unique chance to determine the proto-neutron star radius or the duration of the accretion phase \cite{pagl};  (3) the study of neutral current events will allow us to measure the total amount of energy radiated~\cite{carol}.

\subsection{Extensions of this analysis\label{o35}}

We have adopted 
and validated  
the simplest model to describe the spectrum of the 
neutrino emission, Eq.~\ref{flenza}. However, this  
does not mean that it is impossible  to derive further important information from  SN1987A observations. 

E.g.~a close look at the data allows one to make a point in favor of the existence of a luminous initial  phase of neutrino emission. 
The argument begins by  considering the observations of
 the three detectors in the first second of 
data taking. From Tab.~\ref{tab0}, we see there are 6 events  in   Kamiokande-II; 3 events in IMB; 2 in Baksan: Thus, a large fraction of the whole set of events happened in the first second. 

There are two analyses that have modeled the flux from SN1987A \cite{ll,pagl} and both of them conclude that it is plausible that 
an initial phase of high luminosity occurred. 
Remarkably, this hint corresponds to the expectations of a 
`standard supernova emission', with all caveats discussed in the previous section. 
However, this hint is not very strong;  in \cite{pagl}, e.g.\ it is mentioned that the null hypothesis can be excluded at $\alpha=2$\%.

By contrast, as discussed in Sect.~\ref{p42}, 
it is not clear that a new analysis of the SN1987A data, aimed at 
improving the description of the angular distribution by making allowance for elastic scattering or other non-IBD events, may yield us  useful new information.

In order to  correctly interpret and to take the maximum advantage from  the events that we will collect, in the existing and next generations of neutrino observatories, from a future  galactic supernova, it will be important to implement 
these and other type of extended analyses of the 
neutrino fluxes in future--also accounting for 
the presence of neutrinos of other species and for the possible occurrence of new reactions and signals.
\subsection{Pending issues\label{o4}}
It should be kept in mind that not all issues connected with SN1987A 
have been solved, and a few of them are mentioned here:

1) A major issue is the absence of an observed compact remnant. Not only a pulsar is not seen, but also there is no sign of any point X ray source in the region where the supernova exploded \cite{sht} yet. At the moment, the most plausible conclusion is that we do not understand sufficiently well the cooling of the compact object, but this conclusion does not imply in any manner that we should stop the search for such a remnant which could be the final confirmation of our general understanding of the gravitational collapse (and plausibly an occasion of learning more physics). 
Alternative interpretations include the possible formation of an essentially invisible object (an isolated black hole?), the existence of a dense region of dust that shields the emission, the destruction of the compact remnant after the explosion ...

2) An issue that has been debated in the literature is the meaning of the angular distribution of the  events. Kamiokande-II and IMB data 
point in the direction opposite the supernova more than what we would expect {\em a priori}. However, from a closer 
investigation, one notes that  $(i)$ IMB has an angular bias that favors this outcome, and 
$(ii)$ only Kamiokande-II data has one  event (and just one) that could be attributed to the elastic scattering reaction--see Sect.~\ref{p42}.
Thus, one should be careful before combining the angular distributions of the two experiments. Moreover, the  quasi flat angular distribution that we expect from the IBD hypothesis offers more chances of fluctuations to the meager set of events we are discussing, than the much narrower energy spectrum that has been discussed in this paper. The problem is not very serious on statistical ground \cite{prdML}, and as we have seen it does not significantly affect the inferences.

3) Another issue is the meaning of the 5 events seen by LSD; some tentative interpretations have been proposed after the observation \cite{deruju,bcg,olga}, but they are major modification of a theoretical picture that is already rather incomplete, and this makes it difficult to assess their reliability.  Moreover, the alternative interpretations do not easily reproduce the observed events; the astrophysics involved is quite extreme; the comparison with the observations of the other detectors is  challenging; and it is not easy to imagine convincing reasons for the almost monochromatic spectrum of the 5 events of LSD. In view of these considerations, the events of LSD are usually left aside from the interpretation, and we adhered to such an attitude in the present investigation.

\section{Summary and conclusion\label{o1}}

In this work, we have examined  a number of technical  issues  concerning the analysis of SN1987A events, including a more accurate   description of the cross section, the modeling of the detectors and of their backgrounds, see
Sect.~\ref{pelos}, but we have discussed also the 
assumed form of the fluence, the possibility of extracting more information from the spectrum, etc. We hope that the above discussion will help to summarize and clarify previous results and will prove useful to proceed further. 

The most important scientific conclusion of this study is  simply that the events that have been observed by Kamiokande-II, IMB and Baksan few hours before SN1987A are consistent with the hypothesis that the supernova emitted about $5\times 10^{52}$ erg in electron antineutrinos with an average energy of one dozen of MeV. The errors are not very large: the average energy is known within $\sim 10$\%, while the radiated energy is known within $-20$\% to $+50$\%; the two parameters are correlated by  number of observed events.  See Sect.~\ref{ra} for details.

These conclusions are derived  on the basis of the observation of some tens of events (see Tab.~\ref{tab0}) that according to calculations 
include a bit more than 20 antineutrino signal events. As demonstrated in Sect.~\ref{cc}, the inferences  are not severely affected by any known bias  (e.g.\ inaccurate cross section, procedure of analysis, a priori selection or omission of some data, etc), or incomplete physical description (e.g.\ deviations from thermal distribution, inclusion of neutrino oscillations, etc), namely all the effects introduced in Sects.~\ref{ra} and \ref{aa}. In short, the results are remarkably stable and the analysis provides us with a reliable determination of the couple of parameters that describe the emission. 

Finally, we have argued that there is no clear indication of other physics from the observed energy spectra (Sect.~\ref{p3}) and  
 that the agreement with the theory is satisfactory even though there are some issues that remain unsolved to date (Sect.~\ref{culdis}).



In view of the above considerations, it seems fair to conclude that the observations from SN1987A have been and remain a useful benchmark and a precious occasion to test our knowledge of what happens during a gravitational collapse. Indeed, these data have been used   to  foresee what would be the response of the existing 
detector  to a future supernova (e.g.\  scintillating detectors  \cite{snews,carol}); to plan new detectors; 
 to discuss the possibility of observing relic supernova neutrinos (aka, diffuse supernova background) \cite{aanda};   
to probe neutrino masses \cite{ll,pagl}; 
as an external  trigger to search for gravity wave bursts
\cite{cocc,raffo};~etc.   

%

Certainly,  the observations of neutrinos from a future supernova will allow us to progress greatly in the understanding of the gravitational collapse phenomenon. 
In the meantime, we should get ready for such a epoch-making appointment; in particular, we should be aware of what are 
the borders of the present knowledge,  
be able to use the new observations as effectively as possible,  be prepared for a  comparison with the previous ones.
Actually, a principal motivation of this work is just the hope to contribute to these goals, by clarifying as much as possible the interpretation of SN1987A observations. 

\subsection*{Acknowledgments}
{\footnotesize
I am glad of this occasion to thank my  `supernovae' 
collaborators, namely,    
V.~S.~Berezinsky, 
F.~Cavanna,
M.~Cirelli, 
E.~Coccia, 
M.~L.~Costantini,
A.~Drago,
W.~Fulgione,
P.~L.~Ghia, 
A.~Ianni,
C.~Lujan-Peschard,
G.~Pagliaroli,
F.~Rossi-Torres,
M.~Selvi,
A.~Strumia and
F.~L.~Villante.
Also, I  gratefully acknowledge the help and  
precious conversations with 
E.~N.~Alekseev,  
W.~D.~Arnett, 
G.~Battistoni, 
A.~Bettini,
T.~Cervoni, 
F.~Ferroni, 
P.~Galeotti, 
H.-T.~Janka, 
I.~Krivosheina,
A.~S.~Mal'gin, 
D.~K.~Nadyozhin,
O.~G.~Ryazhskaya, 
O.~Saavedra, 
F.~Scanga, 
A.~Yu.~Smirnov 
and 
C.~Volpe.
I am happy to 
thank G.~G.~Raffelt, S.~Sch$\ddot{\mbox{o}}$nert and 
A.~Yu.~Smirnov for invitation and support at the MIAPP Institute for Astro- and Particle Physics, of the DFG cluster of excellence 
``Origin and Structure of the Universe'',
June-July 2014, where this work has been completed and the results discussed at the NIAPP Topical Workshop on Neutrinos.
Finally, I am grateful to two anonymous Referees of JPhysG for feedback and excellent recommendations.

}



\tableofcontents

\end{document}